\documentclass{aa}  

\usepackage{graphicx}
\usepackage{txfonts}
\usepackage{multirow}
\usepackage[usenames,dvipsnames,svgnames,table]{xcolor}
\usepackage{float}
\usepackage[
breaklinks=true,
hyperindex=true,
colorlinks=true,
linkcolor=Sepia,
citecolor=Purple]{hyperref}
\usepackage{bm}
\usepackage[T1]{fontenc}
\usepackage{tgbonum}
\usepackage{multirow}
\usepackage{booktabs}
\usepackage{arydshln}
\usepackage{subcaption}

\begin{document}

   \title{The interdependence between density PDF, CMF and IMF and their relation with Mach number in simulations}


   \author{A. N\'u\~nez-Casti\~neyra
          \inst{1},
          M. Gonz\'alez \inst{2},
          N. Brucy \inst{3},
          P. Hennebelle \inst{1},
          F. Louvet \inst{4}
          \and
          F. Motte \inst{4}
          \fnmsep
          }

\institute{Université Paris-Saclay, Université Paris Cité, CEA, CNRS, AIM, 91191, Gif-sur-Yvette, France
             \and
             Université Paris Cité, Université Paris-Saclay, CEA, CNRS, AIM, F-91191, Gif-sur-Yvette, France
             \and
             Universität Heidelberg, Zentrum für Astronomie, Institut für Theoretische Astrophysik, Albert-Ueberle-Str 2, D-69120 Heidelberg, Germany
             \and
             Univ. Grenoble Alpes, CNRS, IPAG, 38000 Grenoble, France
             \and
             ...
             }

   \date{submitted December, 2024}

 
  \abstract
   {The initial mass function (IMF) of stars and the corresponding cloud mass function (CMF), traditionally considered universal, exhibit variations that are influenced by the local environment. Notably, these variations are apparent in the distribution's tail, indicating a possible relationship between local dynamics and mass distribution.}
   {Our study is designed to examine how the gas PDF , the IMF and  the CMF depend on the local turbulence within the interstellar medium (ISM).}
   {We run hydrodynamical simulations on small star-forming sections of the ISM under varying turbulence conditions, characterized by Mach numbers of 1, 3.5, and 10, and with two distinct mean densities. This approach allowed us to observe the effects of different turbulence levels on the formation of stellar and cloud masses.}
   {The study demonstrates a clear correlation between the dynamics of the cloud and the IMF. In environments with lower levels of turbulence likely dominated by gravitational collapse, our simulations showed the formation of more massive structures with a powerlaw gas PDF, leading to a top-heavy IMF and CMF. On the other hand environment dominated by turbulence result in a lognormal PDF and a Salpeter-like CMF and IMF. This indicates that the turbulence level is a critical factor in determining the mass distribution within star-forming regions.}
   {}


   \maketitle
%

\section{Introduction}

The stellar initial mass function (IMF), defined as the number density of stars per logarithmic mass interval, $dN/d \log M$ has been the subject of extensive research across various settings, including the Galactic field, young clusters, star-forming regions, as well as the Galactic bulge, halo, and high-redshift galaxies. This research aims to uncover the physical principles shaping its pattern (refer to \citealt{Kroupa2002}, \citealt{Chabrier2003} and \citealt{HennebelleGrudic2024} for comprehensive reviews). Understanding the IMF accurately remains a critical, yet unresolved, challenge in astrophysics. The IMF is crucial for linking stellar and galactic evolution and influences the universe's chemical composition, luminosity, and baryonic content. Additionally, many analyses of the prestellar dense core mass function (CMF) have investigated whether it mirrors the IMF, primarily because it has been proposed that they could present similar forms \citep[see for instance][]{Andre2010,Konyves2015}. In particular, the exponent 
of the power-law, $\alpha$, 
where 
\begin{eqnarray}
     {d n \over d \log m}  \propto m^{ - \alpha +1}.
\end{eqnarray}
that is usually used to describe the stellar mass distribution, has been found to be comparable both for the IMF and for the CMF with typical reported values for $\alpha$ close to 2.35  as originally inferred by \citet{Salpeter1955} although significantly shallower values have also been reported. In the literature, the index may also be found as $\Gamma_{IMF}=-\alpha+1$. 

Another important feature is the peak of the IMF which is usually observed to occur at about 0.3 $M_\odot$. Typically, the CMF is observed to shift by a factor of 2 to 4 towards higher masses relative to the IMF \citep[][]{Motte1998,Testi1998,Johnstone2000, Andre2007,Alves2007,Andre2010,Konyves2015} although \citet{louvet2021} claim that the peak of the CMF that has been reported so far is determined by 
the instrumental resolution. 
For a relation to hold between the CMF and the IMF, there must be a sufficiently accurate correlation between the mass of protostellar cores and the ultimate mass of stars.  However, accurately assessing this correlation continues to be a significant challenge \citep[e.g.][]{smith2009,Ntormousi2019,Smullen2020,Pelkonen2021}.

The shape of the IMF in most studies is derived by constructing a histogram of stellar masses (or their logarithmic values) in equally sized bins, followed by fitting one or several functional forms. Minimizing the chi-square of the fit allows for the derivation of fit parameters. In the intermediate to high-mass regimes, the slopes derived from this method range between 0.7 and 2 when stellar masses are binned logarithmically. Although it is often stated in the literature that the derived values are consistent with the Salpeter slope within the 1$\sigma$ uncertainty, the clarity of this claim is questionable \citep{Dib2014}. An examination of the slope values in this mass regime for several clusters, which employed identical data reduction algorithms and theoretical evolutionary tracks for deriving stellar masses, suggests that the IMF slopes for these clusters do not agree within 1$\sigma$ uncertainty level \citep{Sharma2008,Lata2010,Tripathi2014}. 


\begin{figure*}[h!]
    \centering
    \includegraphics[width=0.98\textwidth]{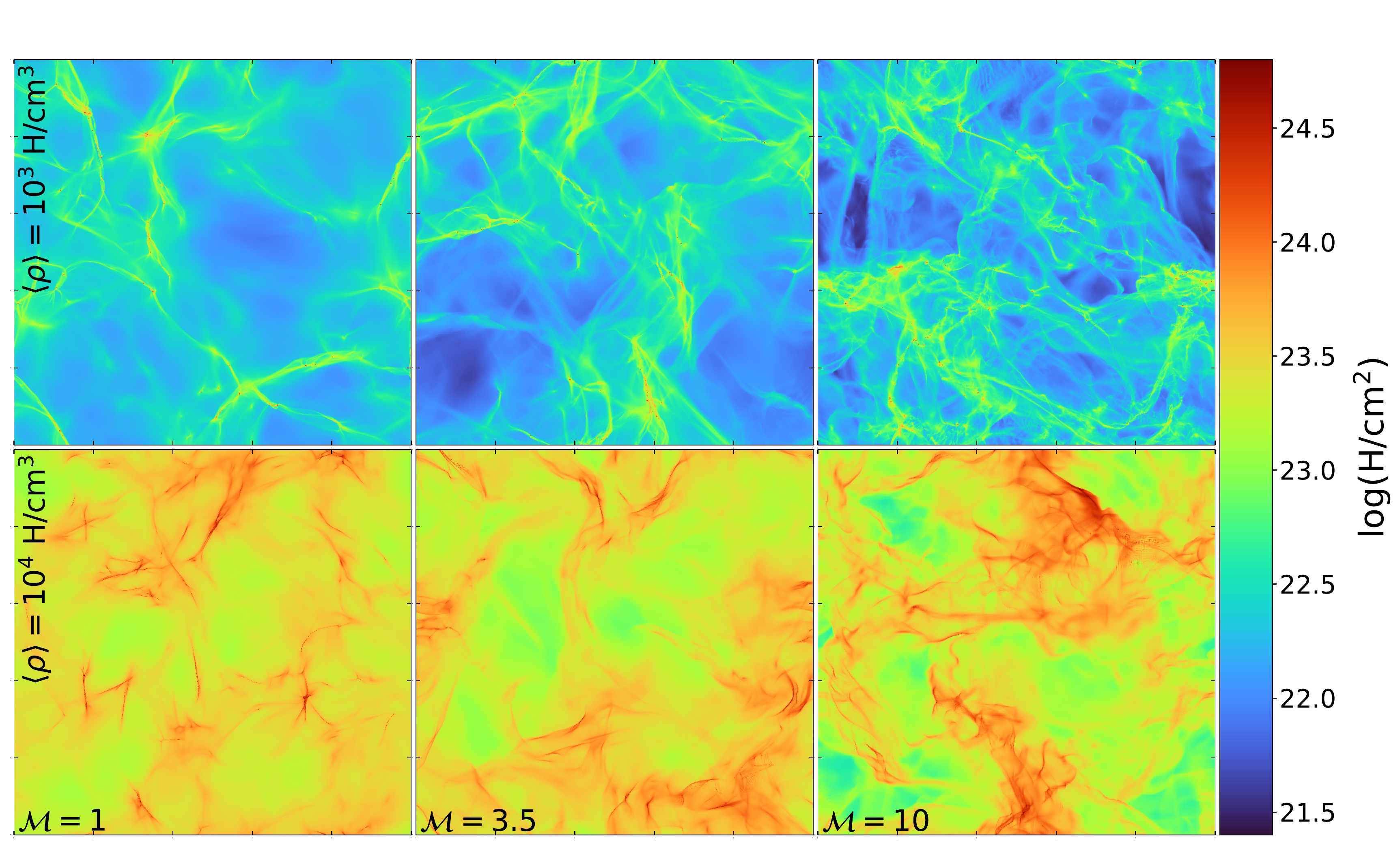}
    \caption{Column density maps for all runs growing in Mach number from left to right and in mean density from top to bottom. This images correspond to the moment when all run have deposit similar amount of mass into stellar particles.}
    \label{fig:mapsrho3}
\end{figure*}

It is becoming increasingly evident that the density structure of the interstellar medium (ISM), where cores form, is predominantly influenced by supersonic turbulence across a broad spectrum of scales \citep{ElmegreenScalo2004,MaclowKlessen2004,McKeeOstriker2007}. A common result of this turbulence is the convergence of the density distribution toward a lognormal probability distribution function (PDF), where the dispersion exhibits a clear dependence on the Mach number (see \citealp{VazquezSemadeni1994,Padoan1997,Scalo1998,Ostriker1999} also see \cite{Hopkins2013} for a more robust PDF description and \cite{Brucy2024} and \cite{Hennebelle2024} for its relation with simulation and star formation theories).

The IMF has been derived from numerical simulations using Lagrangian sink particles \citep[e.g.][]{Krumholz2004,Bleuler2014}. Several studies have been carried out along the years and early calculations include for instance \citet{BateBonnell2005,PadoanNordlund2002,TilleyPudritz2004,Li2004,Ballesteros-Paredes2006,Padoan2007}.

Two main numerical setups are used to compute the IMF, on one-hand isolated turbulent collapsing clumps \citep[e.g.][]{bate2009,ballesteros2015,leeh2018a} and on the other-hand, driven turbulence periodic boxes \citep[e.g.][]{haugbolle2018,mathew2023}. In the former configuration, a turbulent velocity field is initialised and the collapsing clump is then evolving without further driving. \citet{leeh2018a} have investigated the influence of the initial virial  parameter, $\alpha_{vir}$, of the collapsing clump (see their figure 7) between $\alpha_{vir}=0.1$ to 1.5.  They reported  that unless  $\alpha_{vir} < 0.3$,  the influence of $\alpha$ variation remains limited and the slope of the heavy tail of the IMF $\Gamma_{IMF} = -\alpha +1$  remains between -0.8 and -1. The values of $\Gamma_{IMF}$ obtained with turbulent forcing tend to be steeper and closer to $-1.3$. Indeed, the figure~9 of \citet{haugbolle2018} reveals that between $\simeq 2$ and 10 M$_\odot$, the stellar distribution appears to be compatible with $-\alpha+1 \simeq -1.3$.  The work of \citet{mathew2023} presents similar trends and values of $-\alpha +1 \simeq -1.3$ are also inferred (see their figure 7). Recently, \citet{guszejnov2022} have performed a series of numerical simulations corresponding either to periodic boxes with and without turbulent driving or to collapsing clouds. Figure 16 of \citet{guszejnov2022} shows the difference between the various configurations.  All cases have $-\alpha+1 \simeq -1$ except the turbulent driven simulation
for which $-\alpha+1 \simeq -1.3$.

The reason for the differences found when turbulence is driven and when it is not driven, is not elucidated yet. One possibility is that they are a consequence of the density PDF as proposed in \citet{leeh2018a}. Let us remind that in the gravo-turbulent model of \citet{HennebelleChabrier2008} the inferred value of $\alpha$ is such that $\alpha-1 = (n+1)/(2n -4)$ where $n \simeq 3.7$ is the exponent of the velocity power spectrum. To get this relation, a lognormal PDF is assumed. However, when  gravitational collapse is assumed a powerlaw PDF $\propto \rho^{-3/2}$ develops \citep[e.g.][]{kritsuk2011,hf2012}, and the inferred exponent for the IMF is such that $\alpha-1= (5 n - 13) / (2n - 4)$. 

The purpose of the present paper is to better understand the origin of the different regimes that have been inferred for the exponent $\alpha$ and for that purpose we perform a series of high-resolution simulations in which we vary the strength of the turbulent driving, therefore producing flows at various Mach numbers.  In parallel to the IMF, we examine the CMF to evaluate the extent of variation in its exponent and to determine if the exponents of the IMF and CMF are similar and exhibit comparable changes.

In the second part of the paper we present various definitions and discuss how to measure powerlaw exponents. The third part is devoted to the numerical setup. In sections four, five and six, we discuss the density PDF, the CMF and the IMF, respectively. Finally, part seven concludes the paper.

\section{The IMF: definition and slope measurement}

\subsection{Initial mass function }
The IMF as originally defined by \cite{Salpeter1955} is the number of stars $N$ in a volume of space $V$ per logarithmic mass interval $d \log m$:

\begin{equation}\label{eq:logIMF}
    \xi(\log m) = \frac{d(N/V)}{d \log m}= \frac{dn}{d \log m}
\end{equation}

where $n$ is the stellar number density. While this logarithmic form is the most satisfactory representation of the mass distribution in the Galaxy, there is an alternative definition from \cite{Scalo1986} where the mass spectrum is defined in linear mass intervals. It is then clear that:

\begin{equation}\label{eq:linearIMF}
    \xi(m) = \frac{dn}{dm} = \frac{1}{m(\ln 10)} \xi(\log m).
\end{equation}

There are several definitions for the shape of $\xi$, all agreeing on the fact that the tail of the distribution behaves as a power law of the form $\xi(m) \propto m^{-\alpha}$, meaning that $\xi(\log m) \propto m^{-\alpha+1}$. The original value proposed by \cite{Salpeter1955}  was developed on the logarithmic scale and resulted in $\alpha=2.35$\footnote{Note that the index identified in \cite{Salpeter1955} equals 1.35, potentially more recognizable to the reader. Here, $\alpha$ is defined as the index in the linear formulation for the sake of consistency in the analyses that follow, see Table~\ref{tab:indexes}.}  (see \citealp{Chabrier2003} for a review). 

The IMF is represented as a power law probability distribution. This classification is interesting because power law distributions are "heavy-tailed." This means that their distribution of the high-mass elements retains a significant contribution to the total mass.

\begin{figure*}
    \centering
    \includegraphics[width=0.9\textwidth]{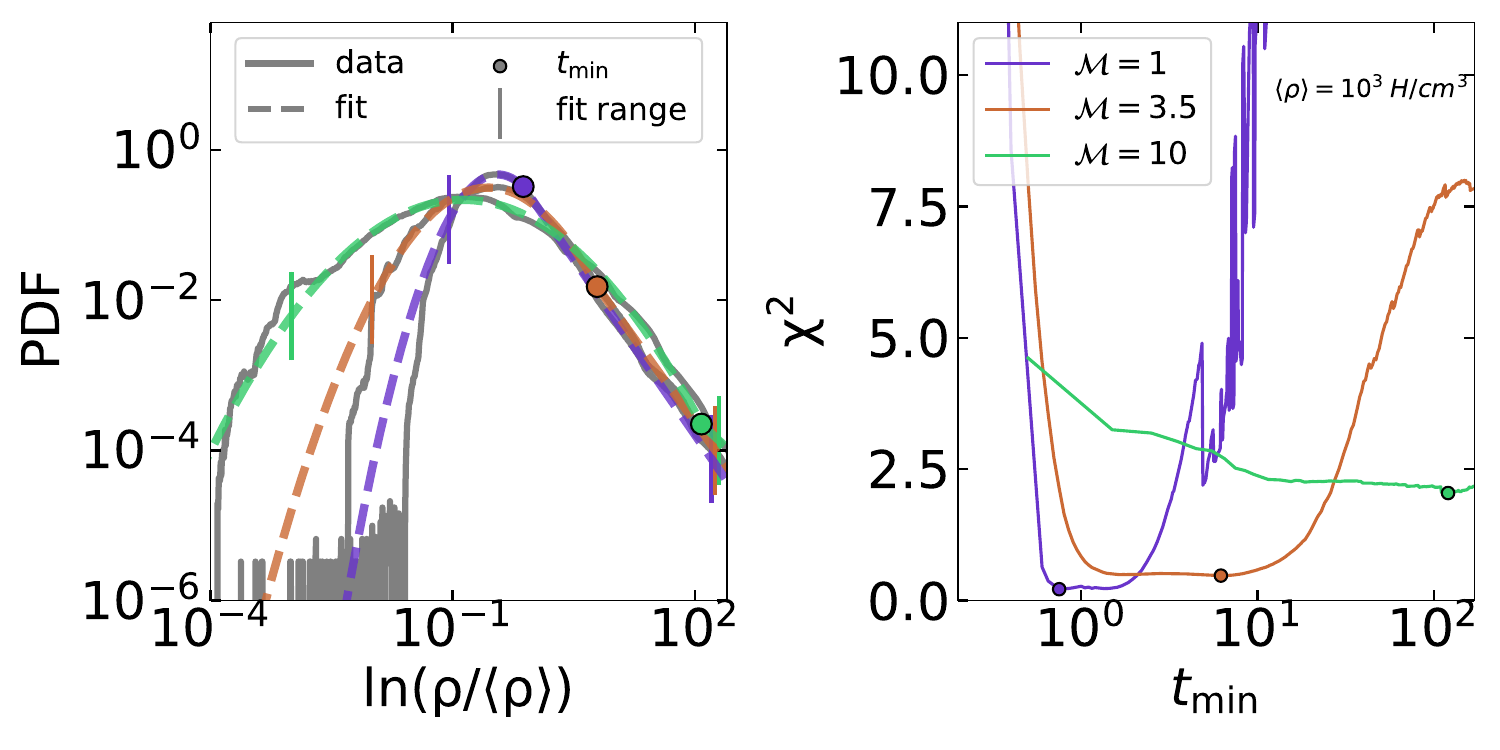}
    \caption{Gas PDF from three simulations with a mean density of $\langle \rho \rangle = 10^3$ H/cm$^3$ (left panel) is depicted. Grey solid lines represent the data, while dashed lines indicate the fits using Equation \ref{eq:fit}. The fitted curves are shown in different colors, representing the three turbulence levels: $\mathcal{M} = $1, 3.5, and 10, displayed in purple, orange, and green, respectively. The circle marker indicates the position of $t_{\rm min}$, while vertical lines mark the start and end of the fit range. The right panel illustrates the relationship between the goodness of fit, measured by the $\chi^2$ value, as $t_{\rm min}$ shifts from the peak to the end of the distribution.}
    \label{fig:PDFrho3}
\end{figure*}

\begin{figure*}
    \centering
    \includegraphics[width=0.9\textwidth]{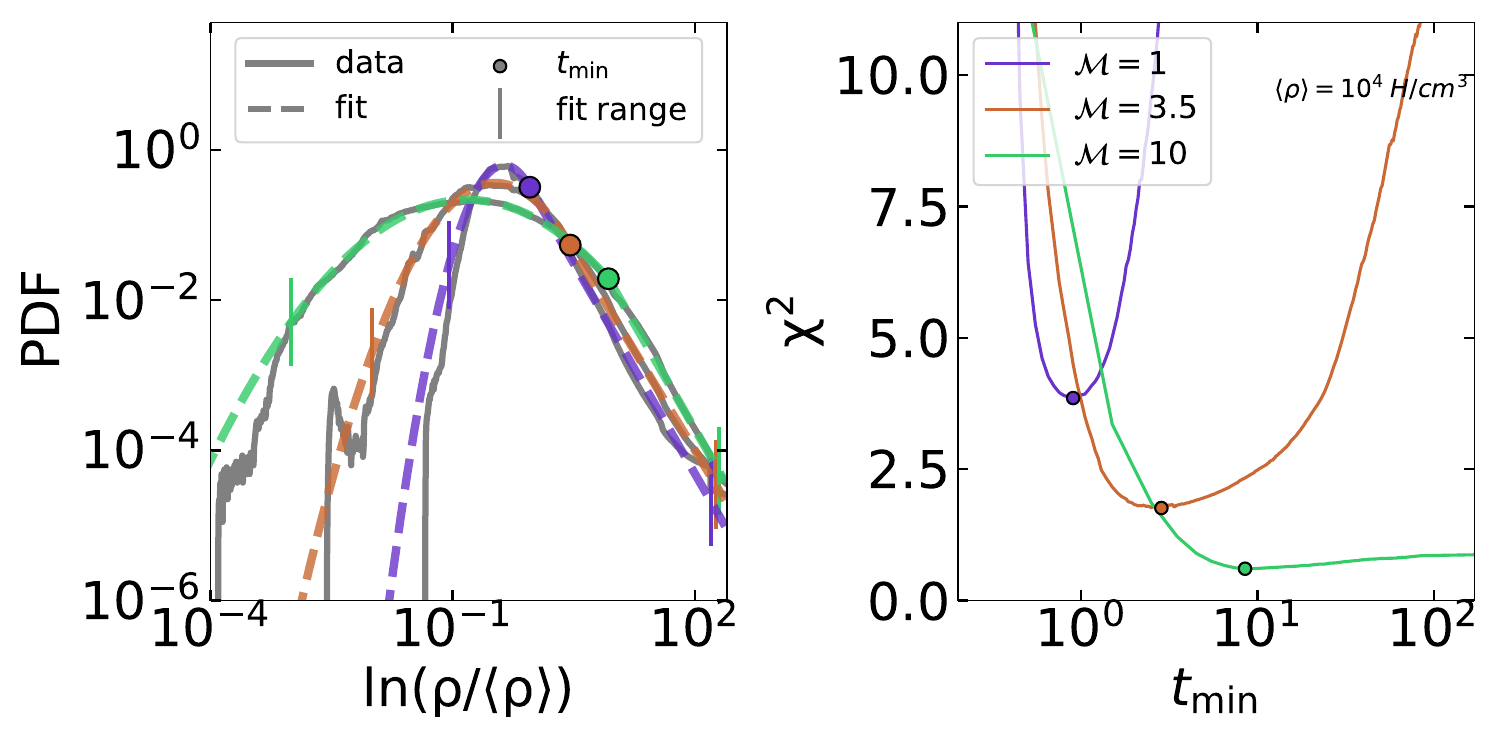}
    \caption{same as in figure \ref{fig:PDFrho3}  but for $\langle \rho \rangle = 10^4$}
    \label{fig:PDFrho4}
\end{figure*}

\begin{figure*}
   \centering
\begin{subfigure}{0.45\linewidth}
       \includegraphics[width=\linewidth]{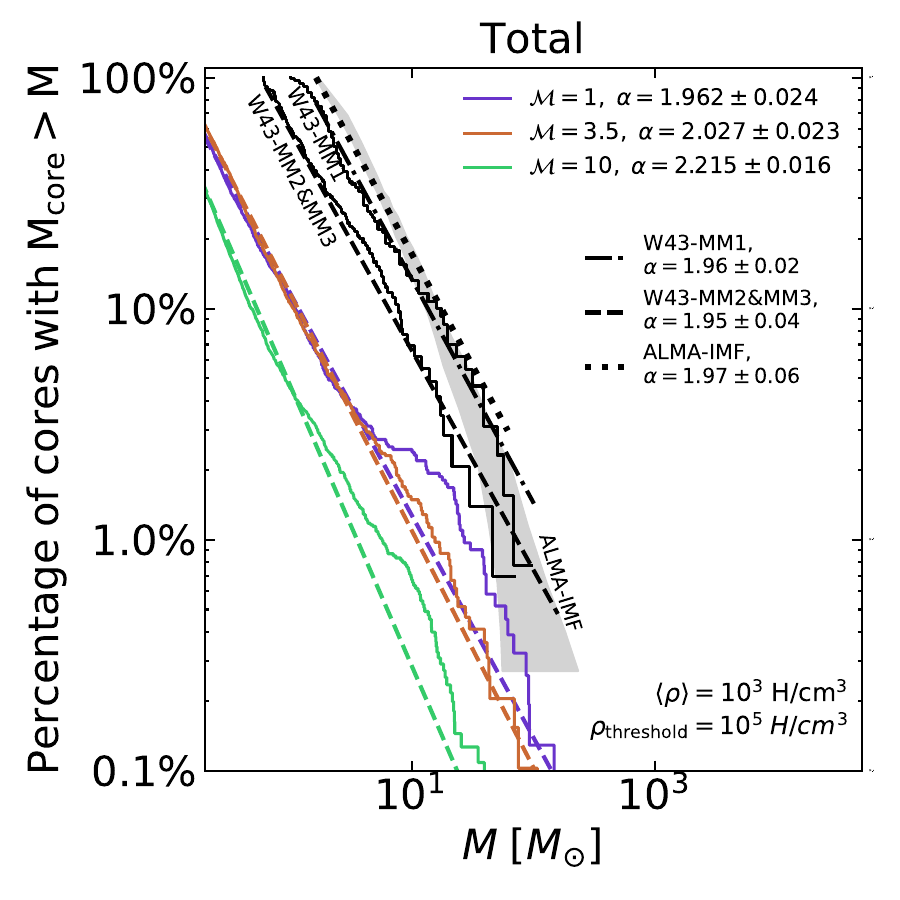}
    \end{subfigure}
\begin{subfigure}{0.45\linewidth}
           \includegraphics[width=\linewidth]{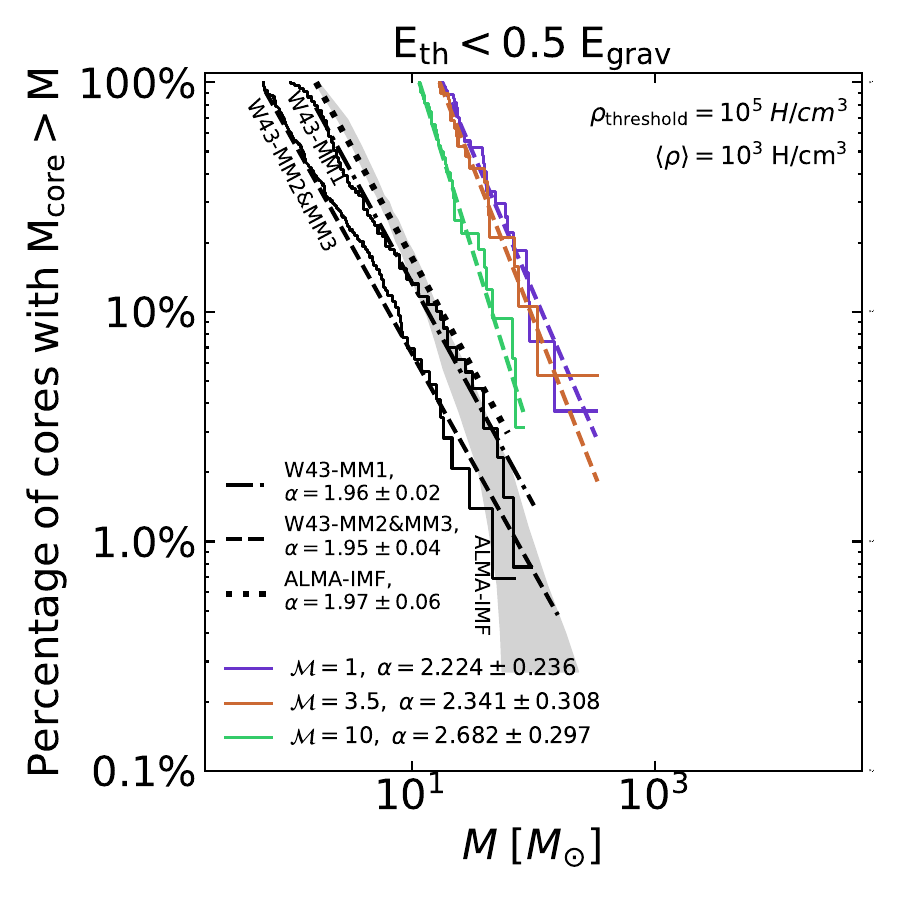}
    \end{subfigure}


\caption{The complementary cumulative distribution function of cores is presented, the total population of cores (left) and those cores meeting the virial condition $E_{\rm th}< 0.5 E_{\rm kin}$ (right), for a mean density of $\langle \rho \rangle = 10^3$ H/cm$^3$. Data for the core populations are depicted with solid lines, whereas estimated slopes for the MLE method are illustrated with dashed lines. Black curves represent observations from ALMA.  }
         \label{fig:CMFtailrho3}
  \end{figure*}

\begin{figure*}
   \centering
\begin{subfigure}{0.45\linewidth}
       \includegraphics[width=\linewidth]{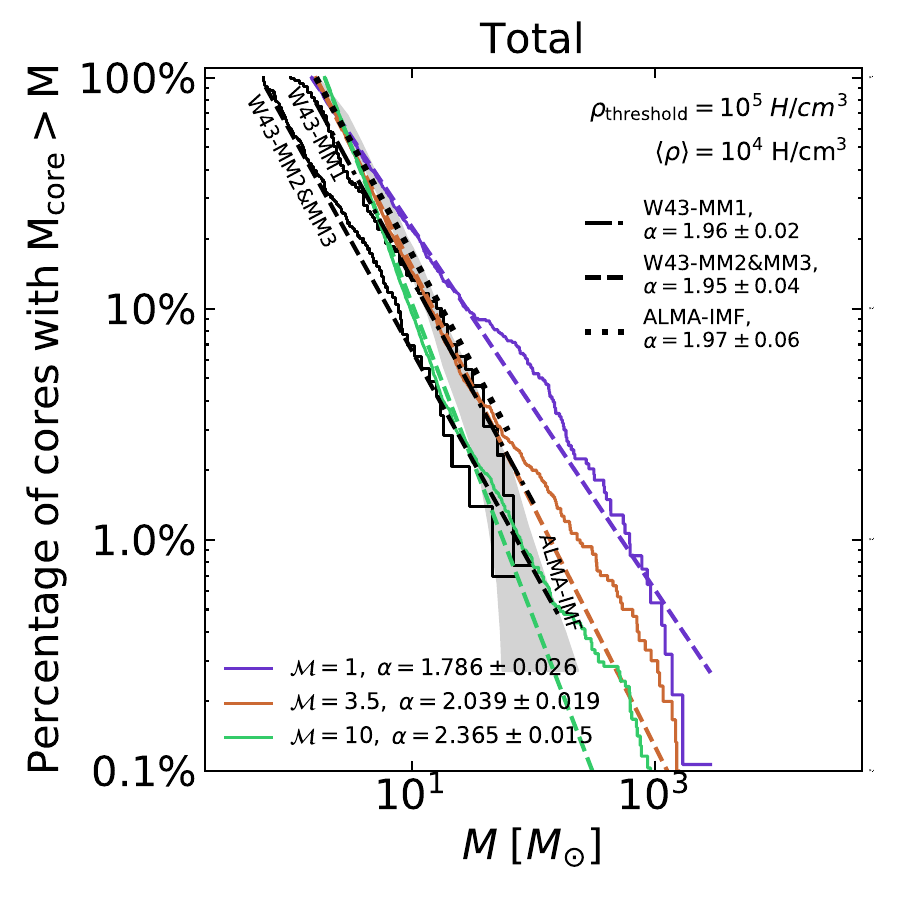}
    \end{subfigure}
\begin{subfigure}{0.45\linewidth}
           \includegraphics[width=\linewidth]{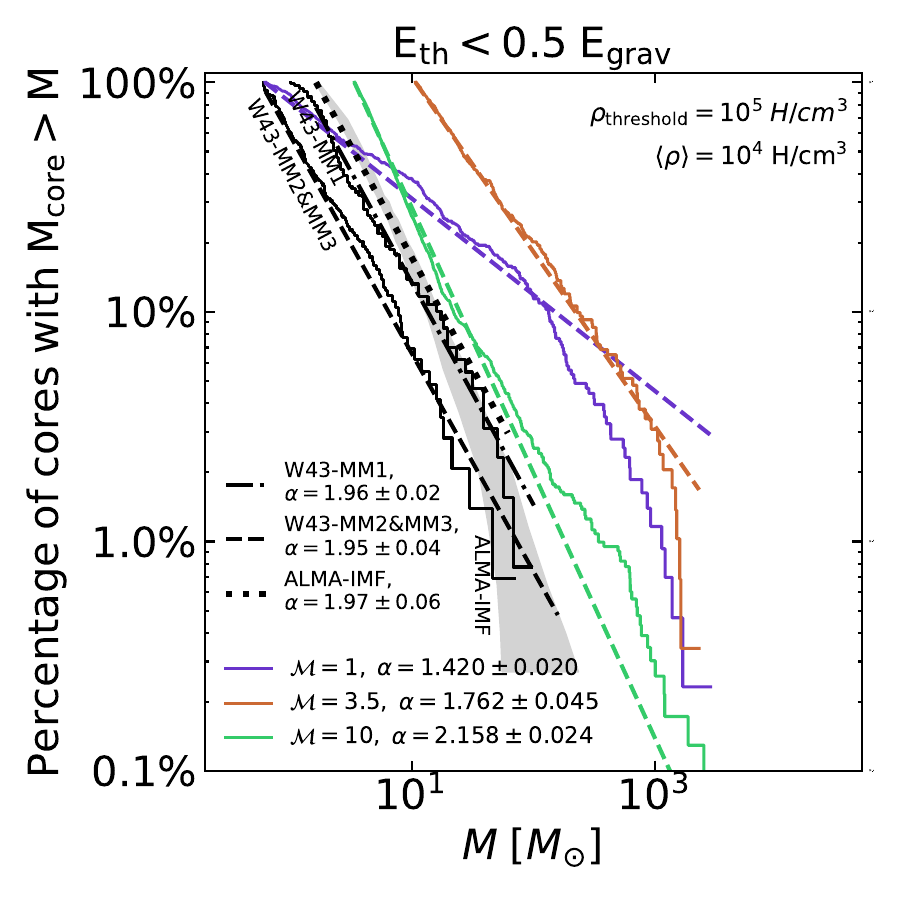}
    \end{subfigure}


\caption{same as in figure \ref{fig:CMFtailrho3}  but for $\langle \rho \rangle = 10^4$ }
         \label{fig:CMFtailrho4}
  \end{figure*}

\subsection{Tail heavy distributions}\label{sec:MLE}

In practice, it is uncommon for empirical phenomena to follow power laws across all values of $x$. Typically, the power law is relevant only for values above a certain minimum $x_{\rm min}$. If we consider the IMF to be a power law probability distribution of the form:

\begin{equation}
    p(x) \textrm{d}x = C x^{-\alpha} \textrm{d}x ,
\end{equation}

where $C$ is a normalization constant. Clearly, this expression diverges as $x \to 0$ therefore it can not describe the probability of all positive values of $x$.  Then a minimum value, $x_{\rm min}$, is required and can be used to compute $C$ assuming $\alpha>1$ as:
\begin{equation}
    p(x)=\frac{\alpha-1}{x_{\rm min}} \left( \frac{x}{x_{\rm min}}\right)^{-\alpha}.
\end{equation}
Once $x_{\rm min}$ is defined, one can estimate the value and respective uncertainty of the power law index $\alpha$ using the Maximum Likelihood Estimate (MLE) method as described in \cite{Clauset2009}. More specifically, using the Python package {\fontfamily{qcr}\selectfont powerlaw} presented in \cite{Alstott2014}. This method provides an accurate estimation of the index $\alpha$ as:

\begin{equation}\label{eq:alphaestimator}
    \hat{\alpha} = 1+n\left[\sum^n_{i
=1} \textrm{ln} \frac{x_i}{x_{\rm min}}  \right],
\end{equation}

where $x_i$ is the mass of the i-th element provided that $x_i > x_{\rm min}$. The uncertainty on $\alpha$ is estimated as:

\begin{equation}
    \hat{\sigma} = \frac{\hat{\alpha} - 1}{\sqrt{n}} + O \left( \frac{1}{n} \right)
    \end{equation}
where $O$ is the mathematical notation for not negligible in front of. 
Opting for this method instead of fitting the tail of the distribution addresses the different issues in this type of analysis. Specifically, representing distributions with histograms followed by linear regression can cause inaccuracies \citep{Clauset2009}. Depicting the probability function in log-log scales doesn't allow for assessing fit uncertainty because of non-Gaussian noise. Furthermore, the bin width selection adds a complicating factor to estimating uncertainty. 

In many cases, it is useful to consider also the complementary cumulative distribution function or CCDF of a power-law distributed variable that can be denoted as:
\begin{equation}
    P(x)= \int^{\infty}_{x} p(x')\textrm{d}x' =\left(\frac{x}{x_{\rm min}}\right)^{-\alpha+1}.
\end{equation}

For distributions such as the CMF and IMF, this method is especially useful because it eliminates the need for binning in plotting, as all that is to be plotted is the exact number of cores of stars above each mass. Thus removing a potential source of uncertainty. However, it introduces a point of confusion: the scaling parameter appears as $-\alpha+1$, coincidentally matching the scaling in the logarithmic version of the IMF from equation \ref{eq:logIMF}, but this similarity is coincidental. For the sake of clarity, Table~\ref{tab:indexes} lists the equations used in upcoming analyses and describes how their power law tails scale. The visual representation of the CCDF is generally more robust against fluctuations due to finite sample sizes, particularly in the tail of the distribution, compared to that of the PDF. Therefore, in the subsequent analysis, the tail of the complementary cumulative distribution function (CCDF), $P(x)$, is shown for cores and stellar particles and not the direct probability density function.

\begin{table}[]
\caption{Different functions to describe the IMF/CMF and the scaling of the tail}

\label{tab:indexes}
\begin{tabular}{|c|l|l|l|}
\hline
                & \multicolumn{1}{c|}{$f(m)$}                                         & \multicolumn{1}{c|}{tail}                           & \multicolumn{1}{c|}{index}       \\ \hline
                &                                                                     &                                                     &                                  \\
linear & \multicolumn{1}{c|}{$\xi(m) = dn/dm$}                               & \multicolumn{1}{c|}{$\propto m^{-\alpha}$}          & \multicolumn{1}{c|}{$-\alpha$}   \\
                &                                                                     &                                                     &                                  \\ \hline
                &                                                                     &                                                     &                                  \\
logaritmic & \multicolumn{1}{c|}{$\xi(\log m) = dn/d \log m $}                   & \multicolumn{1}{c|}{$\propto m \times m^{-\alpha}$} & \multicolumn{1}{c|}{$-\alpha+1$} \\
                &                                                                     &                                                     &                                  \\ \hline
                &                                                                     &                                                     &                                  \\
CCDF            & \multicolumn{1}{c|}{$P(m)= \int^{\infty}_{m} \xi(m')\textrm{d}m' $} & \multicolumn{1}{c|}{$\propto m^{-\alpha+1}$}        & \multicolumn{1}{c|}{$-\alpha+1$} \\
                &                                                                     &                                                     &                                  \\ \hline
\end{tabular}
\tablefoot{The logaritmic version of the mass spectrums and the CCDF have the same scaling for the tail but this is purely coincidental. }
\end{table}

\section{Numerical Setup}
\label{sec:setup}

\subsection{Simulation Framework}
Here we run hydrodynamic simulations for turbulent gas dynamics using the AMR code \textsc{Ramses} \citep{Teyssier2002}. The simulations incorporate a Godunov scheme, the \textsc{hllc} Riemann solver, and \textsc{MinMod} slope limiter.

\subsection{Initial Conditions}
Our simulations commence with a straightforward setup: a uniform-density gas field, driven by large-scale turbulence, in a periodic cubic box. The box, with side lengths of 10 pc reaches a maximum resolution of 63 au. Two sets of runs are carried out at a number densities of $n_0 = 10^3~\mathrm{cm}^{-3}$ or $10^4~\mathrm{cm}^{-3}$. We do not track the gas's chemical evolution and assume a mean molecular mass of $\mu =1.4$ (in atomic mass units $m_p$), maintaining an isothermal state at 10 K. This configuration, influenced by recent studies \citep{Brucy2023}, focuses on how varying initial densities and very high Mach numbers, in the range of hundreds, impact the star formation rate (SFR). 
The simulation is carried out in two steps: first without gravity, and then with gravity. As described in section \ref{subsec:turbinj}, we generate turbulence in the same way during both steps. The main goal is to study how turbulence affects the formation of new stars and self-gravitating gas clouds. In the first step, to make sure we have a realistic turbulent environment, we run the simulation for two turbulence crossing times before turning on gravity. During the second step, we allow stars to form and then examine how they are distributed.

The choices for Mach number and mean density within this very common numerical framework is designed for comparison with ALMA-IMF large program radio observations, which target regions with similar mean densities and turbulence levels. However, it is important to note key differences that fall outside the scope of this exercise. First, the simulations do not include feedback processes, while the observed regions are linked to clusters of massive protostars that eject bipolar outflows observed up to 0.5 pc from their protostellar cores \citep{Nony2020,Nony2024,Towner2024}. Additionally, nearly half of the ALMA-IMF protoclusters are influenced by HII regions, with a moderate to strong impact observed in a quarter of these protoclusters \cite{Motte2022,Galvan-Madrid2024}. Next, these protoclusters are formed by large-scale phenomena that are probably not well represented by the evolution, in a periodic box, of a cloud concentrating under its own gravity and with non-compressive turbulence.
Extracting turbulence levels in a given region is observationally challenging, as it involves distinguishing gas movements caused by gravitational infall from those due to turbulence. This challenge is addressed by focusing on cores or by decomposing the observed gas in less dense clouds into multiple velocity components. In the ALMA-IMF regions, such efforts have yielded Mach numbers around 5 in the cores \cite{Cunningham2023}, using DCN(3-2) which traces gas at densities around $10^7$~cm$^{-3}$. Similarly, Mach numbers peak between 4 and 7, with tails extending up to 20 Koley+ in prep, using the C$^{18}$O(2-1) line which traces gas at densities around $10^3$~cm$^{-3}$. At first order, the presented simulations are suitable for representing these conditions as both, the density and the turbulence level have been chosen to this end


\begin{figure*}
    \centering
        \includegraphics[width=0.95\textwidth]{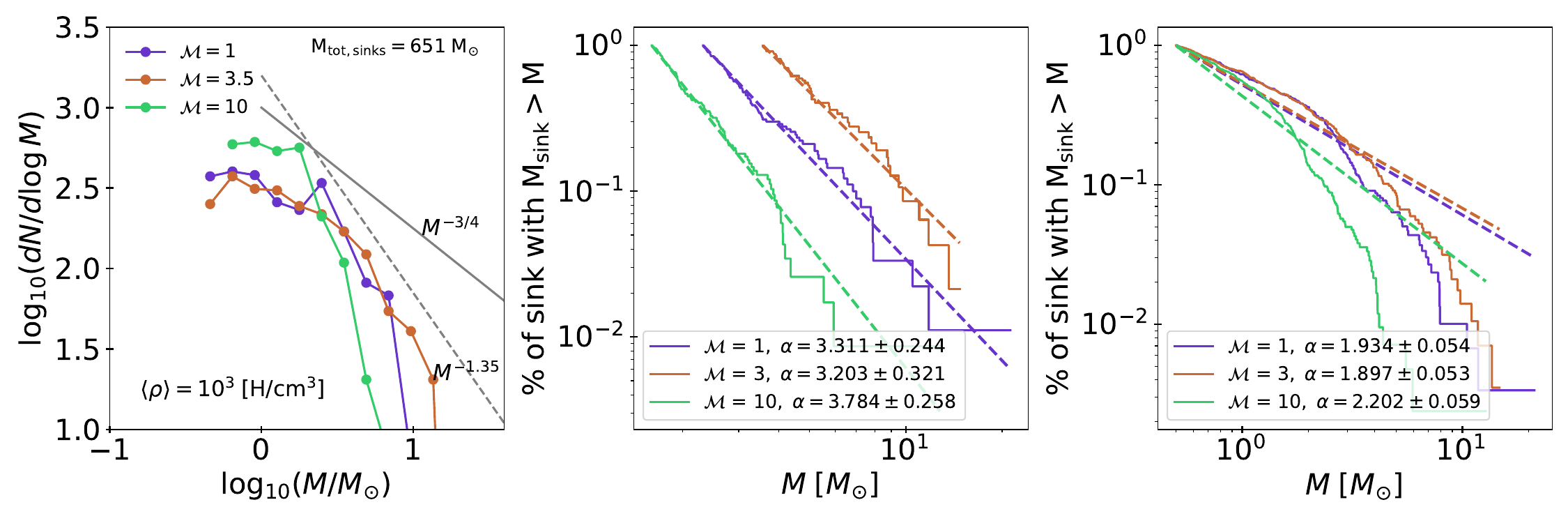}
    \caption{The left panel displays the IMF for simulations conducted at a lower mean density ($\langle \rho \rangle = 10^3$ H/cm$^3$), with dashed and solid lines representing steep (Salpeter) and flatter power law, respectively, for guidance. The middle and right panels present the CCDF, normalized to reflect the percentage of sinks with mass exceeding the values on the horizontal axis. Dashed lines in these panels indicate power laws determined by the Maximum Likelihood Estimate (MLE) method. In the middle panel, the MLE method dynamically estimates the value of $x_{\rm min}$, while in the right panel, $x_{\rm min}$ is set at $0.5$ M$_{\odot}$.}
    \label{fig:IMF_r3}
\end{figure*}

\begin{figure*}
    \centering
    \includegraphics[width=0.95\textwidth]{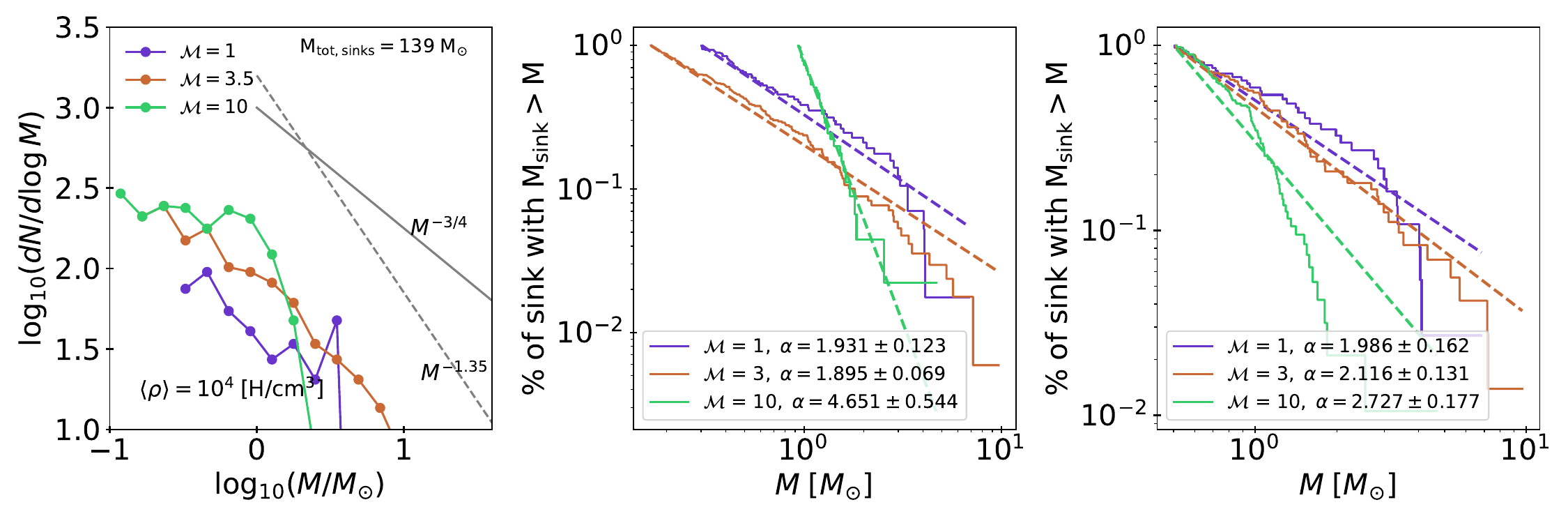}
    \caption{same as figure \ref{fig:IMF_r3}  but for simulations with $\langle \rho \rangle = 10^4$ H/cm$^3$.}
    \label{fig:IMF_r4}
\end{figure*}



\subsection{Turbulence Injection Methodology}
\label{subsec:turbinj}

We use a version of the Ornstein-Uhlenbeck model for turbulence generation \citep{Eswaran1988, Schmidt2006, Schmidt2009, Federrath2010}. The turbulence is continuously injected along the simulations. Here, we outline this model for completeness and to introduce relevant terminology.

We calculate the force driving the turbulence in Fourier space. The evolution of the Fourier modes, $\bm{\hat{f}}$, of said force  are governed by the differential equation:
\begin{equation}
\label{eq:ed_fourier}
    \mathrm{d}\bm{\hat{f}}(\bm{k}, t) = - \bm{\hat{f}}(\bm{k}, t)\dfrac{\mathrm{d}t}{T_\mathrm{driv}} 
    + F_0(\bm{k})\bm{P_\chi}\left(\bm{k}\right) \mathrm{d}\bm{W}_t.
\end{equation}
Here, $T_\mathrm{driv}$ represents the autocorrelation timescale of turbulence, approximately equal to $L_\mathrm{box}/ (2 \sigma)$. Where $\sigma$ is the 3D mass-averaged velocity dispersion computed in the whole simulation. This value divided by the sound speed, $c_{\rm s}$, is used in the definition of the Mach number, $\mathcal{M}=\sigma/c_{\rm s}$. Then  $\mathrm{d}t$ is the integration time step, $\mathrm{d}\bm{W}_t$ is a stochastic term following the Wiener process \citep{Schmidt2009}. The power spectrum of turbulent driving is as follows:
\begin{equation}
    \label{eq:F0}
    F_0(\bm{k}) = 
    \begin{cases} 
    1 - \left(\dfrac{\bm{k}}{2\pi} - 2\right)^2\text{ if } 
    1 < \dfrac{\vert k \vert}{2\pi} < 3 \\
    \;\;\;\; 0 \text{ \;\;\;\;\;\;\;\;
    \;\;\;\;\; if\;not.}
    \end{cases}
\end{equation}
$\bm{P_\chi}(\bm{k})$ is the projection operator that balances compressive and solenoidal modes in the Helmholtz decomposition of one mode versus the other:
\begin{equation}
 \label{eq:projection}
    \bm{P_\chi}(\bm{k}) =  (1 - \chi) \bm{P}^{\perp}(\bm{k}) + 
    \chi \bm{P}^{\parallel}(\bm{k}) 
\end{equation}

with $\bm{P}^{\perp}$ and $\bm{P}^{\parallel}$ being the perpendicular and parallel projection operators with respect to $\bm{k}$ \citep{Federrath2010}. The compressive driving fraction $\chi$ used here is 0 which corresponds to purely solenoidal turbulence.
Finally, the physical force field $\bm{f}(\bm{x}, t)$ in the simulations is derived from the Fourier modes:
\begin{equation}
\label{eq:injection}
\bm{f}(\bm{x}, t) = g(\chi) f_{\mathrm{rms}}  \int\bm{\hat{f}}(\bm{k}, t) 
e^{i\bm{k}\cdot x} d^3\bm{k}
\end{equation}
the coefficient $f_{\mathrm{rms}}$ serves as a proxy to control the energy injected into the simulations by turbulence. The dependence of $\mathcal{M}$ on this coefficient was explored in \citealp{Brucy2024} (see their figure 3). Additionally, $g(\chi)$ is an empirical corrective factor ensuring that the average power across Fourier modes remains consistent with $f_{\mathrm{rms}}$, irrespective of $\chi$. 

\subsection{Self-Gravity and Star Formation}
\label{subsec:gravity}

In the second phase of our simulations, we introduce gravity to assess its impact on the star formation rate (SFR). These simulations commence with the density and velocity fields derived from their respective non-gravitational predecessors at $2 T_\mathrm{driv}$, where $T_\mathrm{driv}$ is the auto-correlation timescale of turbulence, as mentioned in section \ref{subsec:turbinj}. This approach guarantees that turbulence is fully developed before the activation of gravity.

For the calculation of gravitational potential, we use a multigrid Poisson solver. The process of star formation is monitored through the implementation of sink particles, following methodologies such as those described by \cite{Krumholz2004} and \cite{Bleuler2014}. These sink particles are introduced when the gas density surpasses a certain threshold, denoted as $\rho_{\mathrm{sink}}=3 \times 10^{10}$ H cm$^{-3}$.

Figure~\ref{fig:mapsrho3} shows projected column density maps for the runs with $\langle \rho \rangle = 10^3 $ on the top and $10^4$ H/cm$^3$ on the bottom rows, respectively. Both figures have the same scale in the colorbar, this way it is evident how the denser runs differ from the lower-density cases. The Mach number changes from left to right in increasing order as denoted in the bottom part of the lower panels, with $\mathcal{M}$=1, 3.5, 10 in the left, centre and right panels respectively. As turbulence increases, the contrast between the box's densest and lower-density regions becomes more pronounced. Higher turbulence levels create a higher number of dense filaments and empty pockets between them, while lower turbulence results in smoother gas distributions. The simulations presented here use purely solenoidal turbulence, which likely does not reflect all the conditions in the observed region. Future studies shall incorporate mixed turbulence modes. However, it is important to recognize that the lack of compressive turbulence will hinder the formation of dense hubs and ridges, impacting the column density PDF, as suggested by observations. This choice simplifies the current study by reducing the number of variables.

\section{Gas density PDF}
The probability density function (PDF) of gas density in supersonic isothermal flows has been thoroughly examined over time, primarily using numerical simulations.
\cite{VazquezSemadeni1994} and \cite{Nordlund1999} suggested that the gas PDF in these cases follow a lognormal distribution represented as:

\begin{equation}
\mathcal{P}(\delta) = \dfrac{1}{2 \pi \sigma_0^2} \exp\left(-\dfrac{(\delta - \delta_0)^2}{2 \sigma_0^2}\right),
\end{equation}

where $\delta = \ln(\rho / \rho_0)$, $\rho_0$ is the average density, $\sigma_{0}$ is the dispersion,  $\delta_0 = \sigma_0^2 / 2$, and

\begin{equation}
\sigma_0^2 = \ln (1 +b^2\mathcal{M}^2).
\end{equation}

Here, $\mathcal{M}$ is the Mach number and $b$ is an empirical parameter that could take values of $b\simeq 0.5-1$. A transition in the behaviour of the high-density tail of the gas probability density function (PDF) is observed, characterized by a power-law shape in low turbulence cases transitioning to a wider lognormal shape under high turbulence. This evolution remains consistent across varying mean densities. Figures \ref{fig:PDFrho3} and \ref{fig:PDFrho4} display the gas PDF on the right panel, accompanied by their respective fits. To quantify and confirm this transition, a fit was applied to the PDFs, evaluated through a $\chi^2$-test on a function defined in segments:
\begin{equation}\label{eq:fit}
 \ f(x) = \begin{cases} 
          \mathcal{P}(\delta) & \rho\leq t_{\rm min} \\
          C \delta^{\beta} & \rho > t_{\rm min} 
       \end{cases}
    \ .
\end{equation}
Where $t_{\rm min}$ is the transition density above which the PDF behaves as a power law with index $\beta$. The fit is conducted with all parameters left free except for $t_{\rm min}$, which takes values from the peak of the distribution to the high-density end. This test aims to assess the significance of employing a dual-function approach—lognormal for low and mid densities and a power law for high densities—against a singular lognormal function for all densities. The $\chi^2$ values, presented in the right panel of Figures \ref{fig:PDFrho3} and \ref{fig:PDFrho4}, indicate a clear minimum for low turbulence scenarios. This suggests that transitioning from a lognormal to a power law provides not just adequate, but also a precise description of the observed density distribution. For high turbulence scenarios ($\mathcal{M}=10$), the end of the $\chi^2$ curve flattens all the way to the end, signifying that both hypotheses fit the density distributions equally well. However, a singular lognormal description is preferred for its simplicity.

In gravo-turbulent theory, the stellar mass spectrum directly depends on the density PDF. In this context, the final spectrum is sensitive to whether the PDF follows a lognormal function or a power law, with the latter predicting a flatter spectrum tail than the former (see equation 37 and section 5.4 of \citealt{HennebelleGrudic2024}).

\begin{table*}[h]
\centering
\caption{Tail index summary table. Here the power law index of the PDF (third column), Total CMF (fourth), constrained CMF (fifth), total IMF (sixth) and constrained IMF (seventh).}
\label{tab:allsumarised}
\begin{tabular}{|c|c|ccccc|}
\hline
                                &               & \multicolumn{5}{c|}{$\alpha$}                                                                                                                                                                              \\ \cline{3-7} 
$\log_{10}\langle \rho \rangle$ & $\mathcal{M}$ & \multicolumn{1}{c|}{PDF}  & \multicolumn{2}{c|}{CMF}                                                                               & \multicolumn{2}{c|}{IMF}                                              \\ \cline{4-7} 
$\log${[}H/cm$^3${]}            &               & \multicolumn{1}{c|}{}     & \multicolumn{1}{c|}{TOTAL}            & \multicolumn{1}{c|}{E$_{\rm th}$ \textless 0.5 E$_{\rm grav}$} & \multicolumn{1}{c|}{TOTAL}           & M\textgreater{}0.5 M$_{\odot}$ \\ \hline
3                               & 1             & \multicolumn{1}{c|}{1.56} & \multicolumn{1}{c|}{2.362$\pm$ 0.024} & \multicolumn{1}{c|}{2.224$\pm$0.236}                           & \multicolumn{1}{c|}{3.311$\pm$0.244} & 1.934$\pm$0.054                \\ \hline
3                               & 3.5           & \multicolumn{1}{c|}{1.49} & \multicolumn{1}{c|}{2.027$\pm$0.023}  & \multicolumn{1}{c|}{2.341$\pm$0.308}                           & \multicolumn{1}{c|}{3.203$\pm$0.321} & 1.897$\pm$0.053                \\ \hline
3                               & 10            & \multicolumn{1}{c|}{-}    & \multicolumn{1}{c|}{2.215$\pm$0.016}  & \multicolumn{1}{c|}{2.682$\pm$0.297}                           & \multicolumn{1}{c|}{3.784$\pm$0.258} & 2.202$\pm$0.059                \\ \hline
4                               & 1             & \multicolumn{1}{c|}{1.86} & \multicolumn{1}{c|}{1.786$\pm$0.026}  & \multicolumn{1}{c|}{1.420$\pm$0.020}                           & \multicolumn{1}{c|}{1.931$\pm$0.123} & 1.986$\pm$0.162                \\ \hline
4                               & 3.5           & \multicolumn{1}{c|}{1.77} & \multicolumn{1}{c|}{2.039$\pm$0.019}  & \multicolumn{1}{c|}{1.762$\pm$0.045}                           & \multicolumn{1}{c|}{1.895$\pm$0.069} & 2.116$\pm$0.131                \\ \hline
4                               & 10            & \multicolumn{1}{c|}{-}    & \multicolumn{1}{c|}{2.365$\pm$0.015}  & \multicolumn{1}{c|}{2.158$\pm$0.024}                           & \multicolumn{1}{c|}{4.651$\pm$0.544} & 2.727$\pm$0.177                \\ \hline
\end{tabular}
\end{table*}

\section{Cloud mass function and initial mass function}\label{sec:CMF}

Clouds within the simulations are identified using a version of the HOP clumpfinder \citep{Eisenstein1998} called the ecogal wrapper\footnote{The full version is available in the \href{https://gitlab.com/tinecolman/ecogal_tools}{ecogal\_tools} GitLab repository, along with test setups}. While a comprehensive account of the code's strategy and its results on various simulations appears in \cite{Colman2024}, a brief overview is provided here. Initially, the code is provided with a list of cells from an input file, detailing each cell's position (x, y, z coordinates), mass, and density. A search tree based on cell positions facilitates efficient nearest-neighbor identification. Then the code determines the densest neighbour of each cell, within the default number of neighbours, set to 16, following \cite{Eisenstein1998} for SPH simulations. This count is deemed optimal for grid simulations, with a minimum of 4 cells as per \cite{Colman2024}. Increasing the count may yield larger structures but at the expense of calculation time. A `hop' process traces a chain of densest neighbours to group cells around peak density points, forming a structure now considered a clump, which we use as a proxy for stellar cores in the simulation and will be addressed like that from now on. The procedure also involves boundary analysis to detect cells neighbouring different groups. This is done by defining a `saddle density' at these boundaries as the average density between adjacent cells from distinct groups. The highest saddle density between group pairs is stored. For grid simulations, 8 neighbours are considered to maintain consistency with grid regularity. The Ecogal wrapper from \citet{Colman2024}, further evaluates core properties, including size, ellipticity, and internal energies (kinetic, thermal, or gravitational).

In this analysis, a lower threshold for core identification at $10^5$ H/cm$^3$ is applied, regardless of the mean density of the full box. Alternative thresholds were evaluated but deemed unsuitable; lower thresholds identified too many spurious structures as cores, while higher ones led to an under-sampling of cores. However, even with a density threshold not all detected cores are likely to be collapsing structures that will form stars. As an attempt to minimize the inclusion of non-collapsing gas structure, a CMF is constructed only for cores where the internal thermal energy is less than half of their gravitational energy, $E_{\rm th} < 0.5 E_{\rm grav}$, as these will be very likely collapsing cores since gravity dominates. The tail of this constrained CMF and that of the CMF built with the total number of detected cores are shown in figure \ref{fig:CMFtailrho3} for the low-density case and in figure \ref{fig:CMFtailrho4} for the high-density case. This selection criterion focuses on thermal rather than kinetic energy because the calculation of kinetic energy includes infall motions related to gravitational energy, which would incorrectly ignore collapsing cores as dominated by their kinetic energies. The tails of the CCDFs for these CMFs are illustrated in Figures \ref{fig:CMFtailrho3} and \ref{fig:CMFtailrho4}, where the left panels display the total population of detected cores, and the right panels show those meeting the thermal criterion. Solid lines represent the power-law behaviour, with the index $\alpha$ estimated using the MLE method (see section \ref{sec:MLE}). The CCDF of the CMF observed in the W43-MM2\&MM3 ridge by ALMA \citep{Pouteau2022} is depicted in solid black lines including the estimated index.

The W43-MM2\&MM3 ridge, part of the significant W43 molecular cloud complex, includes two primary components: W43-MM2, the second most massive young protocluster in the ALMA-IMF survey and its less massive neighbour, W43-MM3 \citep{Motte2022}. Together, they form a ridge with a total mass of about 3.5$\times 10^4$ M$_{\odot}$ spread over approximately 14 pc$^2$ which corresponds to around 2.5$\times 10^3$ M$_{\odot}/$pc$^2$. This is close to the simulations presented here with mean densities of $10^4$ H/cm$^3$ since the box is of 10 pc side meaning that this corresponds to 2.47$\times 10^3$ M$_{\odot}/$pc$^2$. This ridge is a crucial area of the W43 molecular cloud, located at the intersection of the Scutum-Centaurus spiral arm and the Galactic bar, approximately 5.5~kpc from the Sun. Characterized by its high-density filamentary structures, this region is noted for its efficiency in forming high-mass stars, thus classifying it as a mini-starburst area. The ALMA-IMF consortium reconstructed the CMF in this ridge \citep{Pouteau2022} and found a flatter power law index for the distribution tail with a value of 0.95 on the logarithmic formulation. The correspondent CCDF is shown in figures \ref{fig:CMFtailrho3} and \ref{fig:CMFtailrho4}.

The steepness of the power law tail of the distribution, indicated by $\alpha$, correlates with the Mach number of the ISM in these simulations. A steeper mass spectrum of the cores is observed in scenarios with higher turbulence. This pattern holds whether considering the total population or only those meeting the thermal criterion. In the case of $E_{\rm th} < 0.5 E_{\rm grav}$, which corresponds more closely with the cores chosen by observers, the scenarios with $\mathcal{M} = 3.5$ and $\mathcal{M} = 10$ yield results that compare well with observations from the W43-MM2\&MM3 ridge. The results for the estimation of $\alpha$ for these distributions are summarised in Table~\ref{tab:allsumarised}. Another effect observed here is that, from the total sample to the restricted sample, the CMF tail becomes steeper in low-density cases, whereas the opposite trend is seen in high-density cases.

The IMF generated in the simulations is shown in the left panels of Figures \ref{fig:IMF_r3} and \ref{fig:IMF_r4}, calculated when each simulation has formed the same mass into stars. The central and right panels display the CCDFs of the total stellar populations and the constrained stellar population, respectively. The $\alpha$ estimations in the central panels are done without specifying a value for $x_{\rm min}$, while those in the right panels are estimated with $x_{\rm min} = 0.5$~M$_{\odot}$, as the IMF peak is expected to be around this mass. Each curve generated without a fixed $x_{\rm min}$ starts at the $x_{\rm min}$ value determined by the automated MLE method. Typically, this value corresponds to the beginning of the final section of the curve that behaves like a power law. A specific $x_{\rm min}$ condition is applied because the calculation of $\alpha$ can be skewed by the most massive sink particles. This effect is evident in the central panel, where the unconstrained procedure targets only the high-mass tail, which may show a power-law behavior but does not accurately represent the entire tail. Similar to the results of the CMFs tails, an increase in the steepness of the IMF tail is noted with higher turbulence levels. Nonetheless, values of $\alpha$ computed with an imposed $x_{\rm min}$ of half a solar mass fall within the range of the Salpeter value. All calculated values of $\alpha$ are summarized in Table~\ref{tab:allsumarised}.

\begin{figure}
    \centering
    
    \includegraphics[width=0.45\textwidth]{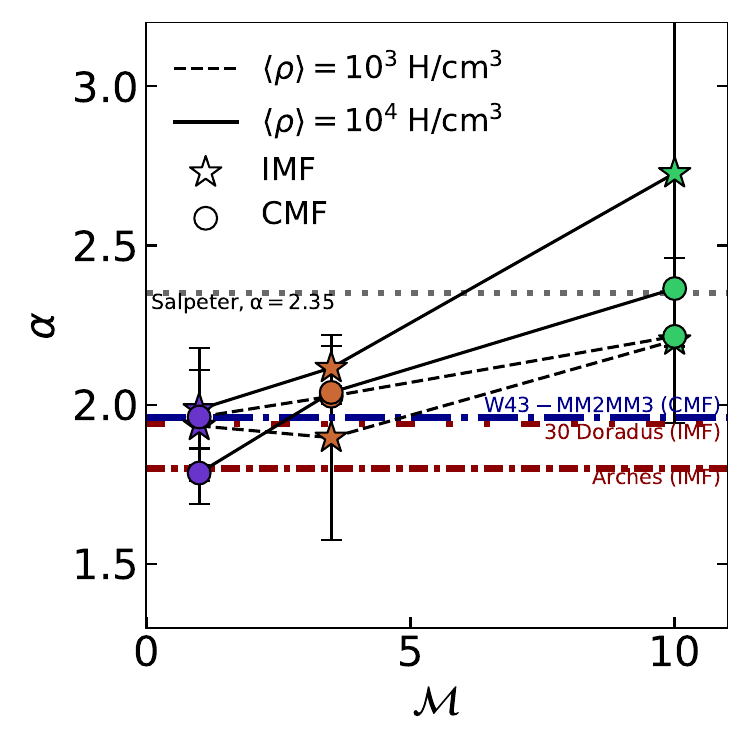}
    \caption{Relation of the power law index $\alpha$ and the Mach number of the different CMF/IMF extracted from the simulations. The stars correspond to the IMF and the circles to the CMF. Dashed lines connect the results from the low-density cases and the solid lines the high density cases. The horizontal grey dotted line shows the Salpeter value, the  blue dot-dashed line is the index corresponding the CMF measured in W43-MM2\&MM3 ridge \citep{Pouteau2022}, the red dot-dot-dashed line corresponds to the index of the IMF tail of 30 Doradus \citep{Schneider2018} and the red dense dot-dashed line shows the index of the IMF tail of the Arches cluster \citep{Hosek2019} .\label{fig:summary}}
\end{figure}

\begin{figure}[h]
    \centering
    
    \includegraphics[width=0.45\textwidth]{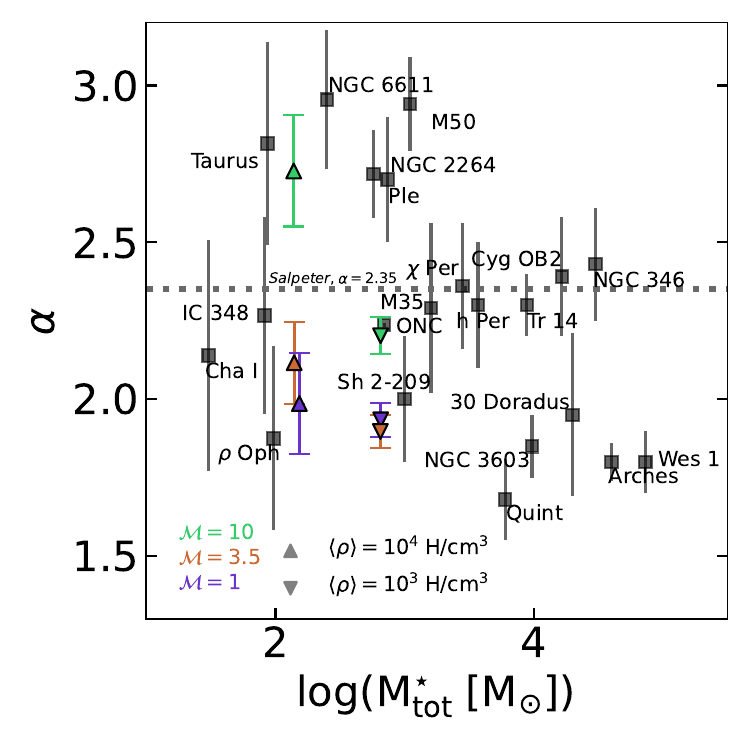}
    \caption{IMF tail power law index $\alpha$ for the high mass tail and its relation with the stellar mass of the cluster for the simulations and obsevations. The horizontal dotted grey line corresponds to the Salpeter value. The values plotted here and their reference can be found in Table~\ref{tab:obs}\label{fig:summaryObs}}
\end{figure}

\section{Discussions}

Figure \ref{fig:summary} shows the power law index, $\alpha$, of the IMF and CMF tail against the Mach number in simulations. For comparison, the horizontal lines show the Salpeter value (dotted gray) and observation results corresponding to the CMF from the W43-MM2\&MM3 ridge (dash dotted blue) and two IMF from massive clusters 30 Doradus (dash-dot-dotted red) and Arches (dense dash-dotted). Population constrains on stars and cores were tested in the simulations and examined in section \ref{sec:CMF}, the summary plot in Figure \ref{fig:summary} shows the results from unconstrained. It is crucial to acknowledge the complexity of star formation physics beyond what is presented here. Not all detected cores may meet the necessary conditions for star formation, and those that do may fragment to form multiple stars. Nonetheless, these findings suggest a connection between the gas PDF, gas core populations, and young stellar populations, consistent with the models proposed by \cite{HennebelleChabrier2008}, particularly in high Mach number scenarios where the gas PDF is lognormal and the CMF/IMF tail approaches a Salpeter-like distribution. The idea is that turbulence works against star formation by providing support against gravitational collapse while also promoting the formation of small structures, therefore, inducing a steep, Salpeter-like tail of the CMF/IMF. It is crucial to recognize that the discussions presented above about turbulence levels, whether high or low, are relevant because all the presented simulations share the same mean density and total mass. It would be more appropriate to focus on whether an environment is dominated by gravity or turbulence. In this context, changes in turbulence levels affect their prominence relative to a common gravitational influence, which remains mostly unchanged between runs. Typically, observed galactic structures are near equipartition or virialization, otherwise total collapse or evaporation would occur. Therefore, as shown here, the degree to which the pressure ratio favours gravity or turbulence will significantly impact star formation. Notably, Figure~\ref{fig:summary} illustrates particular examples of top-heavy CMF from the W43+MM2\&MM3 ridge and IMF from the Arches cluster near the galactic centre and 30 Doradus in the Large Magellanic Cloud. The considerable mass of these clusters suggests that they are predominantly influenced by gravitational collapse rather than by turbulent support, thus, similar to the low-turbulence scenarios discussed here.

Direct comparisons with observations are challenging due to several factors. These simulations are isothermal and scale-independent, whereas the real ISM is neither. On the side of observations, separating velocities from gravitational infall and turbulent gas motions is not trivial. Additionally, while the total mass of young stars in different regions might be similar, the turbulent state of the gas is highly time-dependent and evolves significantly due to feedback. For example, in 30 Doradus and other evolved regions, the gas state reflects the feedback effects and may not correspond to the observed stellar population. As a result, measuring local turbulence directly linked to an IMF is nearly impossible.

One might infer whether a region is gravity-dominated based on its total stellar mass, as massive clusters are more likely to be gravity-dominated. Such a premise is consistent with observations of molecular clouds in the MW, as shown in Figure 17 of \cite{Miville-Deschenes2017}, where the virial parameter decreases with increasing cloud mass, suggesting that more massive clouds are gravity-dominated and prone to collapse. A similar scenario is presented in Figure 1 of \cite{Kauffmann2013} for a more complex sample of star-forming regions. Expanding on this rationale, Figure \ref{fig:summaryObs} illustrates the distribution of the tail slope of the IMF, $\alpha$, and cluster stellar mass for a sample of local and distant clusters. While integrating the simulations presented here into this plot would not provide insightful information as the scale-independent nature of an isothermal box would allow to rescale observables like the total mass, the simulation results are shown as colored errorbar triangle markers. Nevertherless, Figures \ref{fig:summary} and \ref{fig:summaryObs} are complementary, and the underlying rationale remains consistent: observations indicate that  in gravity-dominated regions mostly top-heavy IMFs (low-$\alpha$) are observed, whereas simulations suggest that turbulence-dominated environments result in bottom-heavy CMF/IMF (high, Salpeter-like $\alpha$ and above).
A key feature of this plot is the absence of data in the high-$\alpha$ and high-stellar-mass quadrant, which further supports the proposed rationale. On the left of Figure \ref{fig:summaryObs}, i.e the low-stellar-mass half, clusters consistent with both turbulent-dominated (top-half) and gravity-dominated (bottom-half) scenarios are present. However, mostly clusters consistent with gravity-dominated scenarios are observed in the high-stellar-mass half (the bottom-right quadrant). This trend arises because the turbulent energy required to overcome gravitational collapse in such massive clusters is extremely high and thus rarely achieved.

This picture suggests that proto-stellar cores experience reduced efficiency in accumulating mass through gas accretion within highly turbulent, high-velocity dispersion environments. Notably, \cite{leeh2018a} analytically predicted that a power-law gas PDF produces a flatter stellar mass spectrum with $\alpha \simeq 1.8$, in contrast to a lognormal PDF, which results in Salpeter-like values for $\alpha$, as predicted by \cite{HennebelleChabrier2013}. They argue that, in turbulence-dominated cases, a lognormal gas distribution forms; unlike a power-law PDF, this distribution declines too sharply, limiting the presence of very dense material and thereby reducing the formation of very massive clumps. This prediction aligns with what is observed in the present simulations. The simplicity of these simulations enables, for the first time to our knowledge, a clear link between the gas PDF and the tail of the IMF. For more definitive insights, future studies should include non-isothermal conditions and stellar feedback, as these factors will induce rapid, nonlinear environmental evolution and likely decouple the states of the gas and stellar population.

\section{Conclusions}
Simulations of star-forming regions within the interstellar medium (ISM) were conducted at mean densities of $10^3$ H/cm$^3$ and $10^4$ H/cm$^3$. The isothermal nature of these simulations renders them scale-independent or rescalable. At a spatial scale of 10~pc, the resolution achieved is 64~au. These simulations generate individual star particles, or sink particles, which continue to accrete material under specific conditions. Initially, turbulence is injected in uniformly distributed gas until a predetermined turbulence level is reached, after which gravitational collapse is allowed. The level of turbulence, maintained throughout the simulation, is set at three distinct values of $\mathcal{M}= $1, 3.5, and 10 for each mean density. This setup facilitates the examination of gas properties and the evolution and interaction between the IMF and CMF as turbulence varies. The primary findings are as follows:

\begin{itemize}
\item \textbf{Gas PDF is related to turbulence:} The PDF of the gas density transitions from a power law distribution at lower and medium turbulence levels to a lognormal distribution at the highest Mach number tested, irrespective of the mean density.
\item \textbf{The tail of mass spectra power law of stars and cores evolves with turbulence:} A correlation between the steepness index $\alpha$ of the mass spectrum for cores and stars and the level of turbulence is observed. Scenarios with low turbulence ($\mathcal{M}=1$ and 3.5) exhibit flatter power law indexes for the tails of the CMF and IMF, aligning more closely with observations from galactic star-forming regions as depicted by the dotted line in Figure~\ref{fig:summary}. Conversely, higher turbulence scenarios demonstrate steeper indexes, aligning more closely with the Salpeter slope. This variation, coupled with the fact that the Salpeter slope value is well out of the estimated error bars for low turbulence scenarios in the full populations, challenges the universality of the IMF and underscores the potential for its environmental dependence.
\item \textbf{The CMF and IMF are linked across the tested turbulence range:} The evolution of the slope of the distribution tails, across varying turbulence values, consistently aligns between the CMFs and IMFs within the same simulation. This confirms a connection between the two populations.

\end{itemize}

The present results contribute to addressing several unresolved issues in star formation such as linking the CMF, IMF, and gas PDF, and marking a step toward understanding these phenomena.

\begin{acknowledgements}
      This work has received funding from the French Agence Nationale de la Recherche (ANR) through the project COSMHIC (ANR-20-CE31-0009).
\end{acknowledgements}
\bibliographystyle{aa}
\bibliography{biblio}

\begin{thebibliography}{85}
\expandafter\ifx\csname natexlab\endcsname\relax\def\natexlab#1{#1}\fi

\bibitem[{{Alstott} {et~al.}(2014){Alstott}, {Bullmore}, \& {Plenz}}]{Alstott2014}
{Alstott}, J., {Bullmore}, E., \& {Plenz}, D. 2014, PLoS ONE, 9, e85777

\bibitem[{{Alves} {et~al.}(2007){Alves}, {Lombardi}, \& {Lada}}]{Alves2007}
{Alves}, J., {Lombardi}, M., \& {Lada}, C.~J. 2007, \aap, 462, L17

\bibitem[{{Andersen} {et~al.}(2009){Andersen}, {Zinnecker}, {Moneti}, {McCaughrean}, {Brandl}, {Brandner}, {Meylan}, \& {Hunter}}]{Andersen2009}
{Andersen}, M., {Zinnecker}, H., {Moneti}, A., {et~al.} 2009, \apj, 707, 1347

\bibitem[{{Andr{\'e}} {et~al.}(2007){Andr{\'e}}, {Belloche}, {Motte}, \& {Peretto}}]{Andre2007}
{Andr{\'e}}, P., {Belloche}, A., {Motte}, F., \& {Peretto}, N. 2007, \aap, 472, 519

\bibitem[{{Andr{\'e}} {et~al.}(2010){Andr{\'e}}, {Men'shchikov}, {Bontemps}, {K{\"o}nyves}, {Motte}, {Schneider}, {Didelon}, {Minier}, {Saraceno}, {Ward-Thompson}, {di Francesco}, {White}, {Molinari}, {Testi}, {Abergel}, {Griffin}, {Henning}, {Royer}, {Mer{\'\i}n}, {Vavrek}, {Attard}, {Arzoumanian}, {Wilson}, {Ade}, {Aussel}, {Baluteau}, {Benedettini}, {Bernard}, {Blommaert}, {Cambr{\'e}sy}, {Cox}, {di Giorgio}, {Hargrave}, {Hennemann}, {Huang}, {Kirk}, {Krause}, {Launhardt}, {Leeks}, {Le Pennec}, {Li}, {Martin}, {Maury}, {Olofsson}, {Omont}, {Peretto}, {Pezzuto}, {Prusti}, {Roussel}, {Russeil}, {Sauvage}, {Sibthorpe}, {Sicilia-Aguilar}, {Spinoglio}, {Waelkens}, {Woodcraft}, \& {Zavagno}}]{Andre2010}
{Andr{\'e}}, P., {Men'shchikov}, A., {Bontemps}, S., {et~al.} 2010, \aap, 518, L102

\bibitem[{{Ascenso} {et~al.}(2007){Ascenso}, {Alves}, {Vicente}, \& {Lago}}]{Ascenso2007}
{Ascenso}, J., {Alves}, J., {Vicente}, S., \& {Lago}, M.~T.~V.~T. 2007, \aap, 476, 199

\bibitem[{{Ballesteros-Paredes} {et~al.}(2006){Ballesteros-Paredes}, {Gazol}, {Kim}, {Klessen}, {Jappsen}, \& {Tejero}}]{Ballesteros-Paredes2006}
{Ballesteros-Paredes}, J., {Gazol}, A., {Kim}, J., {et~al.} 2006, \apj, 637, 384

\bibitem[{{Ballesteros-Paredes} {et~al.}(2015){Ballesteros-Paredes}, {Hartmann}, {P{\'e}rez-Goytia}, \& {Kuznetsova}}]{ballesteros2015}
{Ballesteros-Paredes}, J., {Hartmann}, L.~W., {P{\'e}rez-Goytia}, N., \& {Kuznetsova}, A. 2015, \mnras, 452, 566

\bibitem[{{Bate}(2009)}]{bate2009}
{Bate}, M.~R. 2009, \mnras, 392, 590

\bibitem[{{Bate} \& {Bonnell}(2005)}]{BateBonnell2005}
{Bate}, M.~R. \& {Bonnell}, I.~A. 2005, \mnras, 356, 1201

\bibitem[{Bleuler \& Teyssier(2014)}]{Bleuler2014}
Bleuler, A. \& Teyssier, R. 2014, MNRAS, 445, 4015

\bibitem[{{Brucy} {et~al.}(2023){Brucy}, {Hennebelle}, {Colman}, \& {Iteanu}}]{Brucy2023}
{Brucy}, N., {Hennebelle}, P., {Colman}, T., \& {Iteanu}, S. 2023, \aap, 675, A144

\bibitem[{{Brucy} {et~al.}(2024){Brucy}, {Hennebelle}, {Colman}, {Klessen}, \& {Le Yhuelic}}]{Brucy2024}
{Brucy}, N., {Hennebelle}, P., {Colman}, T., {Klessen}, R.~S., \& {Le Yhuelic}, C. 2024, arXiv e-prints, arXiv:2404.17374

\bibitem[{{Chabrier}(2003)}]{Chabrier2003}
{Chabrier}, G. 2003, \pasp, 115, 763

\bibitem[{{Clauset} {et~al.}(2009){Clauset}, {Shalizi}, \& {Newman}}]{Clauset2009}
{Clauset}, A., {Shalizi}, C.~R., \& {Newman}, M.~E.~J. 2009, SIAM Review, 51, 661

\bibitem[{{Colman} {et~al.}(2024){Colman}, {Brucy}, {Girichidis}, {Glover}, {Benedettini}, {Soler}, {Tress}, {Traficante}, {Hennebelle}, {Klessen}, {Molinari}, \& {Miville-Desch{\^e}nes}}]{Colman2024}
{Colman}, T., {Brucy}, N., {Girichidis}, P., {et~al.} 2024, arXiv e-prints, arXiv:2403.00512

\bibitem[{{Cunningham} {et~al.}(2023){Cunningham}, {Ginsburg}, {Galv{\'a}n-Madrid}, {Motte}, {Csengeri}, {Stutz}, {Fern{\'a}ndez-L{\'o}pez}, {{\'A}lvarez-Guti{\'e}rrez}, {Armante}, {Baug}, {Bonfand}, {Bontemps}, {Braine}, {Brouillet}, {Busquet}, {D{\'\i}az-Gonz{\'a}lez}, {Di Francesco}, {Gusdorf}, {Herpin}, {Liu}, {L{\'o}pez-Sepulcre}, {Louvet}, {Lu}, {Maud}, {Nony}, {Olguin}, {Pouteau}, {Rivera-Soto}, {Sandoval-Garrido}, {Sanhueza}, {Tatematsu}, {Towner}, \& {Valeille-Manet}}]{Cunningham2023}
{Cunningham}, N., {Ginsburg}, A., {Galv{\'a}n-Madrid}, R., {et~al.} 2023, \aap, 678, A194

\bibitem[{{Dib}(2014)}]{Dib2014}
{Dib}, S. 2014, \mnras, 444, 1957

\bibitem[{{Eisenstein} \& {Hut}(1998)}]{Eisenstein1998}
{Eisenstein}, D.~J. \& {Hut}, P. 1998, \apj, 498, 137

\bibitem[{{Elmegreen} \& {Falgarone}(1996)}]{ElmegreenFalgarone1996}
{Elmegreen}, B.~G. \& {Falgarone}, E. 1996, \apj, 471, 816

\bibitem[{{Elmegreen} \& {Scalo}(2004)}]{ElmegreenScalo2004}
{Elmegreen}, B.~G. \& {Scalo}, J. 2004, \araa, 42, 211

\bibitem[{Eswaran \& Pope(1988)}]{Eswaran1988}
Eswaran, V. \& Pope, S.~B. 1988, Computers and Fluids, 16, 257

\bibitem[{{Federrath} {et~al.}(2010){Federrath}, {Roman-Duval}, {Klessen}, {Schmidt}, \& {Mac Low}}]{Federrath2010}
{Federrath}, C., {Roman-Duval}, J., {Klessen}, R.~S., {Schmidt}, W., \& {Mac Low}, M.~M. 2010, \aap, 512, A81

\bibitem[{{Galv{\'a}n-Madrid} {et~al.}(2024){Galv{\'a}n-Madrid}, {D{\'\i}az-Gonz{\'a}lez}, {Motte}, {Ginsburg}, {Cunningham}, {Menten}, {Armante}, {Bonfand}, {Braine}, {Csengeri}, {Dell'Ova}, {Louvet}, {Nony}, {Rivera-Soto}, {Sanhueza}, {Stutz}, {Wyrowski}, {{\'A}lvarez-Guti{\'e}rrez}, {Baug}, {Bontemps}, {Bronfman}, {Fern{\'a}ndez-L{\'o}pez}, {Gusdorf}, {Koley}, {Liu}, {Salinas}, {Towner}, \& {Whitworth}}]{Galvan-Madrid2024}
{Galv{\'a}n-Madrid}, R., {D{\'\i}az-Gonz{\'a}lez}, D.~J., {Motte}, F., {et~al.} 2024, \apjs, 274, 15

\bibitem[{{Guszejnov} {et~al.}(2022){Guszejnov}, {Grudi{\'c}}, {Offner}, {Faucher-Gigu{\`e}re}, {Hopkins}, \& {Rosen}}]{guszejnov2022}
{Guszejnov}, D., {Grudi{\'c}}, M.~Y., {Offner}, S. S.~R., {et~al.} 2022, \mnras, 515, 4929

\bibitem[{{Harayama} {et~al.}(2008){Harayama}, {Eisenhauer}, \& {Martins}}]{Harayama2008}
{Harayama}, Y., {Eisenhauer}, F., \& {Martins}, F. 2008, \apj, 675, 1319

\bibitem[{{Haugb{\o}lle} {et~al.}(2018){Haugb{\o}lle}, {Padoan}, \& {Nordlund}}]{haugbolle2018}
{Haugb{\o}lle}, T., {Padoan}, P., \& {Nordlund}, {\r{A}}. 2018, \apj, 854, 35

\bibitem[{{Hennebelle} {et~al.}(2024){Hennebelle}, {Brucy}, \& {Colman}}]{Hennebelle2024}
{Hennebelle}, P., {Brucy}, N., \& {Colman}, T. 2024, \aap, 690, A43

\bibitem[{{Hennebelle} \& {Chabrier}(2008)}]{HennebelleChabrier2008}
{Hennebelle}, P. \& {Chabrier}, G. 2008, \apj, 684, 395

\bibitem[{{Hennebelle} \& {Chabrier}(2013)}]{HennebelleChabrier2013}
{Hennebelle}, P. \& {Chabrier}, G. 2013, \apj, 770, 150

\bibitem[{{Hennebelle} \& {Falgarone}(2012)}]{hf2012}
{Hennebelle}, P. \& {Falgarone}, E. 2012, \aapr, 20, 55

\bibitem[{{Hennebelle} \& {Grudi{\'c}}(2024)}]{HennebelleGrudic2024}
{Hennebelle}, P. \& {Grudi{\'c}}, M.~Y. 2024, arXiv e-prints, arXiv:2404.07301

\bibitem[{{Hopkins}(2013)}]{Hopkins2013}
{Hopkins}, P.~F. 2013, \mnras, 430, 1880

\bibitem[{{Hosek} {et~al.}(2019){Hosek}, {Lu}, {Anderson}, {Najarro}, {Ghez}, {Morris}, {Clarkson}, \& {Albers}}]{Hosek2019}
{Hosek}, Matthew~W., J., {Lu}, J.~R., {Anderson}, J., {et~al.} 2019, \apj, 870, 44

\bibitem[{{Hur} {et~al.}(2012){Hur}, {Sung}, \& {Bessell}}]{Hur2012}
{Hur}, H., {Sung}, H., \& {Bessell}, M.~S. 2012, \aj, 143, 41

\bibitem[{{Hu{\ss}mann} {et~al.}(2012){Hu{\ss}mann}, {Stolte}, {Brandner}, {Gennaro}, \& {Liermann}}]{Hussmann2012}
{Hu{\ss}mann}, B., {Stolte}, A., {Brandner}, W., {Gennaro}, M., \& {Liermann}, A. 2012, \aap, 540, A57

\bibitem[{{Johnstone} {et~al.}(2000){Johnstone}, {Wilson}, {Moriarty-Schieven}, {Joncas}, {Smith}, {Gregersen}, \& {Fich}}]{Johnstone2000}
{Johnstone}, D., {Wilson}, C.~D., {Moriarty-Schieven}, G., {et~al.} 2000, \apj, 545, 327

\bibitem[{{Kalirai} {et~al.}(2003){Kalirai}, {Fahlman}, {Richer}, \& {Ventura}}]{Kalirai2003}
{Kalirai}, J.~S., {Fahlman}, G.~G., {Richer}, H.~B., \& {Ventura}, P. 2003, \aj, 126, 1402

\bibitem[{{Kauffmann} {et~al.}(2013){Kauffmann}, {Pillai}, \& {Goldsmith}}]{Kauffmann2013}
{Kauffmann}, J., {Pillai}, T., \& {Goldsmith}, P.~F. 2013, \apj, 779, 185

\bibitem[{{K{\"o}nyves} {et~al.}(2015){K{\"o}nyves}, {Andr{\'e}}, {Men'shchikov}, {Palmeirim}, {Arzoumanian}, {Schneider}, {Roy}, {Didelon}, {Maury}, {Shimajiri}, {Di Francesco}, {Bontemps}, {Peretto}, {Benedettini}, {Bernard}, {Elia}, {Griffin}, {Hill}, {Kirk}, {Ladjelate}, {Marsh}, {Martin}, {Motte}, {Nguy{\^e}n Luong}, {Pezzuto}, {Roussel}, {Rygl}, {Sadavoy}, {Schisano}, {Spinoglio}, {Ward-Thompson}, \& {White}}]{Konyves2015}
{K{\"o}nyves}, V., {Andr{\'e}}, P., {Men'shchikov}, A., {et~al.} 2015, \aap, 584, A91

\bibitem[{{Kritsuk} {et~al.}(2011){Kritsuk}, {Norman}, \& {Wagner}}]{kritsuk2011}
{Kritsuk}, A.~G., {Norman}, M.~L., \& {Wagner}, R. 2011, \apjl, 727, L20

\bibitem[{{Kroupa}(2002)}]{Kroupa2002}
{Kroupa}, P. 2002, Science, 295, 82

\bibitem[{{Krumholz} {et~al.}(2004){Krumholz}, {McKee}, \& {Klein}}]{Krumholz2004}
{Krumholz}, M.~R., {McKee}, C.~F., \& {Klein}, R.~I. 2004, \apj, 611, 399

\bibitem[{{Lata} {et~al.}(2010){Lata}, {Pandey}, {Kumar}, {Bhatt}, {Pace}, \& {Sharma}}]{Lata2010}
{Lata}, S., {Pandey}, A.~K., {Kumar}, B., {et~al.} 2010, \aj, 139, 378

\bibitem[{{Lee} \& {Hennebelle}(2018)}]{leeh2018a}
{Lee}, Y.-N. \& {Hennebelle}, P. 2018, \aap, 611, A88

\bibitem[{{Li} {et~al.}(2004){Li}, {Klessen}, \& {Mac Low}}]{Li2004}
{Li}, Y., {Klessen}, R.~S., \& {Mac Low}, M.-M. 2004, Baltic Astronomy, 13, 377

\bibitem[{{Lim} {et~al.}(2013){Lim}, {Chun}, {Sung}, {Park}, {Lee}, {Sohn}, {Hur}, \& {Bessell}}]{Lim2013}
{Lim}, B., {Chun}, M.-Y., {Sung}, H., {et~al.} 2013, \aj, 145, 46

\bibitem[{{Louvet} {et~al.}(2021){Louvet}, {Hennebelle}, {Men'shchikov}, {Didelon}, {Ntormousi}, \& {Motte}}]{louvet2021}
{Louvet}, F., {Hennebelle}, P., {Men'shchikov}, A., {et~al.} 2021, \aap, 653, A157

\bibitem[{{Mac Low} \& {Klessen}(2004)}]{MaclowKlessen2004}
{Mac Low}, M.-M. \& {Klessen}, R.~S. 2004, Reviews of Modern Physics, 76, 125

\bibitem[{{Mathew} {et~al.}(2023){Mathew}, {Federrath}, \& {Seta}}]{mathew2023}
{Mathew}, S.~S., {Federrath}, C., \& {Seta}, A. 2023, \mnras, 518, 5190

\bibitem[{{McKee} \& {Ostriker}(2007)}]{McKeeOstriker2007}
{McKee}, C.~F. \& {Ostriker}, E.~C. 2007, \araa, 45, 565

\bibitem[{{Miville-Desch{\^e}nes} {et~al.}(2017){Miville-Desch{\^e}nes}, {Murray}, \& {Lee}}]{Miville-Deschenes2017}
{Miville-Desch{\^e}nes}, M.-A., {Murray}, N., \& {Lee}, E.~J. 2017, \apj, 834, 57

\bibitem[{{Moraux} {et~al.}(2004){Moraux}, {Kroupa}, \& {Bouvier}}]{Moraux2004}
{Moraux}, E., {Kroupa}, P., \& {Bouvier}, J. 2004, \aap, 426, 75

\bibitem[{{Motte} {et~al.}(1998){Motte}, {Andre}, \& {Neri}}]{Motte1998}
{Motte}, F., {Andre}, P., \& {Neri}, R. 1998, \aap, 336, 150

\bibitem[{{Motte} {et~al.}(2022){Motte}, {Bontemps}, {Csengeri}, {Pouteau}, {Louvet}, {Stutz}, {Cunningham}, {L{\'o}pez-Sepulcre}, {Brouillet}, {Galv{\'a}n-Madrid}, {Ginsburg}, {Maud}, {Men'shchikov}, {Nakamura}, {Nony}, {Sanhueza}, {{\'A}lvarez-Guti{\'e}rrez}, {Armante}, {Baug}, {Bonfand}, {Busquet}, {Chapillon}, {D{\'\i}az-Gonz{\'a}lez}, {Fern{\'a}ndez-L{\'o}pez}, {Guzm{\'a}n}, {Herpin}, {Liu}, {Olguin}, {Towner}, {Bally}, {Battersby}, {Braine}, {Bronfman}, {Chen}, {Dell'Ova}, {Di Francesco}, {Gonz{\'a}lez}, {Gusdorf}, {Hennebelle}, {Izumi}, {Joncour}, {Lee}, {Lefloch}, {Lesaffre}, {Lu}, {Menten}, {Mignon-Risse}, {Molet}, {Moraux}, {Mundy}, {Nguyen Luong}, {Reyes}, {Reyes Reyes}, {Robitaille}, {Rosolowsky}, {Sandoval-Garrido}, {Schuller}, {Svoboda}, {Tatematsu}, {Thomasson}, {Walker}, {Wu}, {Whitworth}, \& {Wyrowski}}]{Motte2022}
{Motte}, F., {Bontemps}, S., {Csengeri}, T., {et~al.} 2022, \aap, 662, A8

\bibitem[{{Nony} {et~al.}(2024){Nony}, {Galv{\'a}n-Madrid}, {Brouillet}, {Su{\'a}rez}, {Louvet}, {De Pree}, {Ju{\'a}rez-Gama}, {Ginsburg}, {Immer}, {Lin}, {Liu}, {Rom{\'a}n-Z{\'u}{\~n}iga}, \& {Zhang}}]{Nony2024}
{Nony}, T., {Galv{\'a}n-Madrid}, R., {Brouillet}, N., {et~al.} 2024, \aap, 687, A84

\bibitem[{{Nony} {et~al.}(2020){Nony}, {Motte}, {Louvet}, {Plunkett}, {Gusdorf}, {Fechtenbaum}, {Pouteau}, {Lefloch}, {Bontemps}, {Molet}, \& {Robitaille}}]{Nony2020}
{Nony}, T., {Motte}, F., {Louvet}, F., {et~al.} 2020, \aap, 636, A38

\bibitem[{{Nordlund} \& {Padoan}(1999)}]{Nordlund1999}
{Nordlund}, {\r{A}}.~K. \& {Padoan}, P. 1999, in Interstellar Turbulence, ed. J.~{Franco} \& A.~{Carraminana}, 218

\bibitem[{{Ntormousi} \& {Hennebelle}(2019)}]{Ntormousi2019}
{Ntormousi}, E. \& {Hennebelle}, P. 2019, \aap, 625, A82

\bibitem[{{Ostriker} {et~al.}(1999){Ostriker}, {Gammie}, \& {Stone}}]{Ostriker1999}
{Ostriker}, E.~C., {Gammie}, C.~F., \& {Stone}, J.~M. 1999, \apj, 513, 259

\bibitem[{{Padoan} {et~al.}(1997){Padoan}, {Jones}, \& {Nordlund}}]{Padoan1997}
{Padoan}, P., {Jones}, B. J.~T., \& {Nordlund}, {\r{A}}.~P. 1997, \apj, 474, 730

\bibitem[{{Padoan} \& {Nordlund}(2002)}]{PadoanNordlund2002}
{Padoan}, P. \& {Nordlund}, {\r{A}}. 2002, \apj, 576, 870

\bibitem[{{Padoan} {et~al.}(2007){Padoan}, {Nordlund}, {Kritsuk}, {Norman}, \& {Li}}]{Padoan2007}
{Padoan}, P., {Nordlund}, {\r{A}}., {Kritsuk}, A.~G., {Norman}, M.~L., \& {Li}, P.~S. 2007, \apj, 661, 972

\bibitem[{{Pelkonen} {et~al.}(2021){Pelkonen}, {Padoan}, {Haugb{\o}lle}, \& {Nordlund}}]{Pelkonen2021}
{Pelkonen}, V.~M., {Padoan}, P., {Haugb{\o}lle}, T., \& {Nordlund}, {\r{A}}. 2021, \mnras, 504, 1219

\bibitem[{{Pouteau} {et~al.}(2022){Pouteau}, {Motte}, {Nony}, {Galv{\'a}n-Madrid}, {Men'shchikov}, {Bontemps}, {Robitaille}, {Louvet}, {Ginsburg}, {Herpin}, {L{\'o}pez-Sepulcre}, {Dell'Ova}, {Gusdorf}, {Sanhueza}, {Stutz}, {Brouillet}, {Thomasson}, {Armante}, {Baug}, {Bonfand}, {Busquet}, {Csengeri}, {Cunningham}, {Fern{\'a}ndez-L{\'o}pez}, {Liu}, {Olguin}, {Towner}, {Bally}, {Braine}, {Bronfman}, {Joncour}, {Gonz{\'a}lez}, {Hennebelle}, {Lu}, {Menten}, {Moraux}, {Tatematsu}, {Walker}, \& {Whitworth}}]{Pouteau2022}
{Pouteau}, Y., {Motte}, F., {Nony}, T., {et~al.} 2022, \aap, 664, A26

\bibitem[{{Sabbi} {et~al.}(2008){Sabbi}, {Sirianni}, {Nota}, {Tosi}, {Gallagher}, {Smith}, {Angeretti}, {Meixner}, {Oey}, {Walterbos}, \& {Pasquali}}]{Sabbi2008}
{Sabbi}, E., {Sirianni}, M., {Nota}, A., {et~al.} 2008, \aj, 135, 173

\bibitem[{{Salpeter}(1955)}]{Salpeter1955}
{Salpeter}, E.~E. 1955, \apj, 121, 161

\bibitem[{{Sanhueza} {et~al.}(2019){Sanhueza}, {Contreras}, {Wu}, {Jackson}, {Guzm{\'a}n}, {Zhang}, {Li}, {Lu}, {Silva}, {Izumi}, {Liu}, {Miura}, {Tatematsu}, {Sakai}, {Beuther}, {Garay}, {Ohashi}, {Saito}, {Nakamura}, {Saigo}, {Veena}, {Nguyen-Luong}, \& {Tafoya}}]{Sanhueza2019}
{Sanhueza}, P., {Contreras}, Y., {Wu}, B., {et~al.} 2019, \apj, 886, 102

\bibitem[{{Scalo} {et~al.}(1998){Scalo}, {V{\'a}zquez-Semadeni}, {Chappell}, \& {Passot}}]{Scalo1998}
{Scalo}, J., {V{\'a}zquez-Semadeni}, E., {Chappell}, D., \& {Passot}, T. 1998, \apj, 504, 835

\bibitem[{{Scalo}(1986)}]{Scalo1986}
{Scalo}, J.~M. 1986, \fcp, 11, 1

\bibitem[{Schmidt {et~al.}(2009)Schmidt, Federrath, Hupp, Kern, \& Niemeyer}]{Schmidt2009}
Schmidt, W., Federrath, C., Hupp, M., Kern, S., \& Niemeyer, J.~C. 2009, A\&A, 494, 127

\bibitem[{Schmidt {et~al.}(2006)Schmidt, Hillebrandt, \& Niemeyer}]{Schmidt2006}
Schmidt, W., Hillebrandt, W., \& Niemeyer, J.~C. 2006, Computers \& Fluids, 35, 353

\bibitem[{{Schneider} {et~al.}(2018){Schneider}, {Sana}, {Evans}, {Bestenlehner}, {Castro}, {Fossati}, {Gr{\"a}fener}, {Langer}, {Ram{\'\i}rez-Agudelo}, {Sab{\'\i}n-Sanjuli{\'a}n}, {Sim{\'o}n-D{\'\i}az}, {Tramper}, {Crowther}, {de Koter}, {de Mink}, {Dufton}, {Garcia}, {Gieles}, {H{\'e}nault-Brunet}, {Herrero}, {Izzard}, {Kalari}, {Lennon}, {Ma{\'\i}z Apell{\'a}niz}, {Markova}, {Najarro}, {Podsiadlowski}, {Puls}, {Taylor}, {van Loon}, {Vink}, \& {Norman}}]{Schneider2018}
{Schneider}, F.~R.~N., {Sana}, H., {Evans}, C.~J., {et~al.} 2018, Science, 359, 69

\bibitem[{{Sharma} {et~al.}(2008){Sharma}, {Pandey}, {Ogura}, {Aoki}, {Pandey}, {Sandhu}, \& {Sagar}}]{Sharma2008}
{Sharma}, S., {Pandey}, A.~K., {Ogura}, K., {et~al.} 2008, \aj, 135, 1934

\bibitem[{{Slesnick} {et~al.}(2002){Slesnick}, {Hillenbrand}, \& {Massey}}]{Slesnick2002}
{Slesnick}, C.~L., {Hillenbrand}, L.~A., \& {Massey}, P. 2002, \apj, 576, 880

\bibitem[{{Smith} {et~al.}(2009){Smith}, {Clark}, \& {Bonnell}}]{smith2009}
{Smith}, R.~J., {Clark}, P.~C., \& {Bonnell}, I.~A. 2009, \mnras, 396, 830

\bibitem[{{Smullen} {et~al.}(2020){Smullen}, {Kratter}, {Offner}, {Lee}, \& {Chen}}]{Smullen2020}
{Smullen}, R.~A., {Kratter}, K.~M., {Offner}, S. S.~R., {Lee}, A.~T., \& {Chen}, H. H.-H. 2020, \mnras, 497, 4517

\bibitem[{{Testi} \& {Sargent}(1998)}]{Testi1998}
{Testi}, L. \& {Sargent}, A.~I. 1998, \apjl, 508, L91

\bibitem[{{Teyssier}(2002)}]{Teyssier2002}
{Teyssier}, R. 2002, \aap, 385, 337

\bibitem[{{Tilley} \& {Pudritz}(2004)}]{TilleyPudritz2004}
{Tilley}, D.~A. \& {Pudritz}, R.~E. 2004, \mnras, 353, 769

\bibitem[{{Towner} {et~al.}(2024){Towner}, {Ginsburg}, {Dell'Ova}, {Gusdorf}, {Bontemps}, {Csengeri}, {Galv{\'a}n-Madrid}, {Louvet}, {Motte}, {Sanhueza}, {Stutz}, {Bally}, {Baug}, {Chen}, {Cunningham}, {Fern{\'a}ndez-L{\'o}pez}, {Liu}, {Lu}, {Nony}, {Valeille-Manet}, {Wu}, {{\'A}lvarez-Guti{\'e}rrez}, {Bonfand}, {Di Francesco}, {Nguyen-Luong}, {Olguin}, \& {Whitworth}}]{Towner2024}
{Towner}, A.~P.~M., {Ginsburg}, A., {Dell'Ova}, P., {et~al.} 2024, \apj, 960, 48

\bibitem[{{Tripathi} {et~al.}(2014){Tripathi}, {Pandey}, \& {Kumar}}]{Tripathi2014}
{Tripathi}, A., {Pandey}, U.~S., \& {Kumar}, B. 2014, \na, 29, 1

\bibitem[{{Vazquez-Semadeni}(1994)}]{VazquezSemadeni1994}
{Vazquez-Semadeni}, E. 1994, \apj, 423, 681

\bibitem[{{Wright} {et~al.}(2015){Wright}, {Drew}, \& {Mohr-Smith}}]{Wright2015}
{Wright}, N.~J., {Drew}, J.~E., \& {Mohr-Smith}, M. 2015, \mnras, 449, 741

\bibitem[{{Yasui} {et~al.}(2023){Yasui}, {Kobayashi}, {Saito}, {Izumi}, \& {Ikeda}}]{Yasui2023}
{Yasui}, C., {Kobayashi}, N., {Saito}, M., {Izumi}, N., \& {Ikeda}, Y. 2023, \apj, 943, 137

\end{thebibliography}

\begin{appendix} 

\section{Mass size relation:}
\begin{figure}[h]
    \centering
    
    \includegraphics[width=0.45\textwidth]{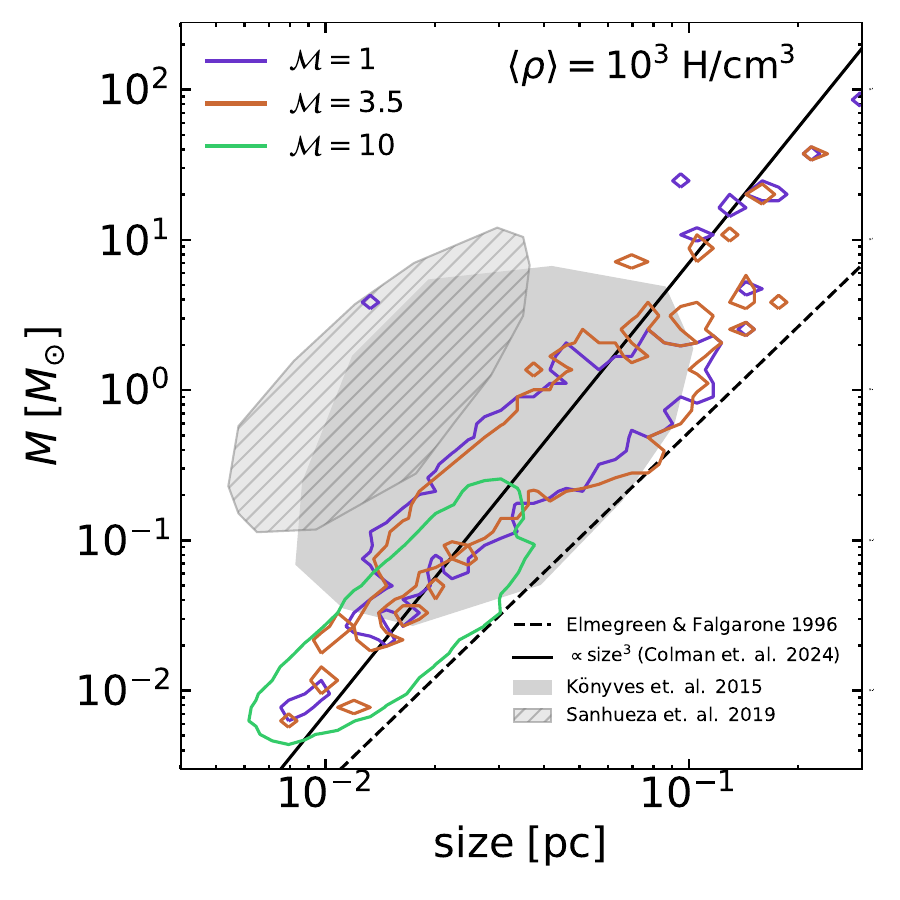}
    \caption{ The mass versus size distribution of the recovered cores from the simulation with a mean density of $10^3$~H/cm$^3$ is shown in comparison with the core samples of \cite{Konyves2015} (solid gray shaded region) and \cite{Sanhueza2019} (gray hatched region). The contours cover 90$\%$ of the recovered cores from the simulations. The dashed line represents the mass-size relation from \cite{ElmegreenFalgarone1996}, and the solid line illustrates the power-law behavior observed in simulations across different scales by \cite{Colman2024}.\label{fig:MassSizerho3}}
\end{figure}

\begin{figure}[h]
    \centering
    
    \includegraphics[width=0.45\textwidth]{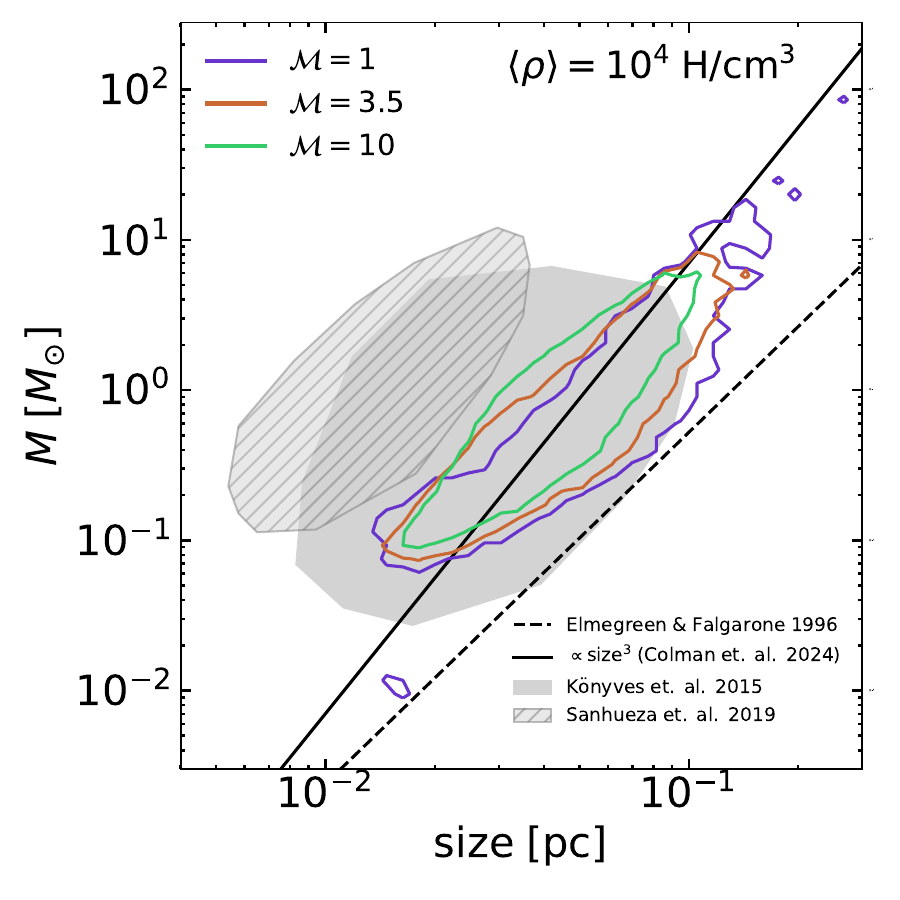}
    \caption{ Same as in figure \ref{fig:MassSizerho3}  but for simulations with $\langle \rho \rangle = 10^4$ H/cm$^3$ \label{fig:MassSizerho4} }
\end{figure}

Figures \ref{fig:MassSizerho3} and \ref{fig:MassSizerho4} display the mass-size relation of cores from different simulations and their comparison with various observations. The simulation results align relatively well with observations of the Aquila region in the Gould Belt \cite{Konyves2015}, a low star-forming region with a quiescent environment. In contrast, ASHES clumps \cite{Sanhueza2019}, associated with high-mass star formation, exhibit higher densities, pressures, and turbulent energies, reflecting an environment that has evolved further due to stellar activity. As a result, these clumps are more massive and do not compare well with the simulations presented here. 
Additionally, the recovered mass-size relation is fully consistent with the results of \cite{Colman2024}, which demonstrated that core mass scales as size$^3$ (black solid line) in simulations across various scales. Here, size is defined as the mean length of the three principal axes of an ellipsoid enclosing all the core’s mass.
Runs with low mean density form smaller and less massive halos as turbulence increases. This leads to a higher total number of cores, consequently affecting the uncertainties in the slope derivations of the CMFs computed in section~\ref{sec:CMF}

\begin{table*}[p]
\caption{Observation of the high mass tail of the stellar initial mass function of different stellar cluster and their stellar mass. These values are plotted in figure \ref{fig:summaryObs}}
\label{tab:obs}
\begin{tabular}{|c|c|c|c|c|}
\hline
name               & $\alpha$          & $M_{*}${[}M\_\{\textbackslash{}odot\}{]} & year & Reference                                             \\ \hline
Taurus             & 1.815 $\pm$ 0.324 & 87.0                                     & 2014 &                                                       \\
IC 348             & 1.267 $\pm$ 0.313 & 82.22                                    & 2014 &                                                       \\
NGC 6611           & 1.954 $\pm$ 0.221 & 250.3                                    & 2014 &                                                       \\
NGC 2264           & 1.717 $\pm$ 0.14  & 574.8                                    & 2014 & \cite{Dib2014}                       \\
ONC                & 1.236 $\pm$ 0.007 & 693.5                                    & 2014 &                                                       \\
$\rho$ Oph         & 0.875 $\pm$ 0.293 & 96.49                                    & 2014 &                                                       \\
Cha I              & 1.139 $\pm$ 0.367 & 30.42                                    & 2014 &                                                       \\ \cline{5-5} 
NGC 3603           & 0.85 $\pm$ 0.1    & 9700.0                                   & 2008 & \cite{Harayama2008}                  \\ \cline{5-5} 
30 Doradus         & 0.95 $\pm$ 0.26   & 4$\times 10 ^4$                          & 2018 & \cite{Schneider2018,Andersen2009}    \\ \cline{5-5} 
Arches             & 0.8 $\pm$ 0.06    & 2$\times 10 ^4$                          & 2018 & \cite{Hosek2019}                     \\ \cline{5-5} 
Cyg OB2            & 1.39 $\pm$ 0.19   & 1.65$\times 10^4$                                    & 2015 & \cite{Wright2015}                    \\ \cline{5-5} 
h Per              & 1.3 $\pm$ 0.2     & 3700                                     & 2002 & \multirow{2}{*}{\cite{Slesnick2002}} \\
$\chi$ Per         & 1.36 $\pm$ 0.20   & 2800                                     & 2002 &                                                       \\ \cline{5-5} 
M35                & 1.29 $\pm$ 0.27   & 1600                                     & 2003 & \multirow{2}{*}{\cite{Kalirai2003}}  \\
M50                & 1.94 $\pm$ 0.15   & 1100                                     & 2003 &                                                       \\ \cline{5-5} 
NGC 346  & 1.43$\pm$ 0.18              & 3$\times 10^4$                                    & 2007 & \cite{Sabbi2008}                     \\ \cline{5-5} 
Ple                & 1.7 $\pm$ 0.2     & 740                                      & 2004 & \cite{Moraux2004}                    \\ \cline{5-5} 
Quint              & 0.68 $\pm$ 0.13   & 6000                                     & 2012 & \cite{Hussmann2012}                  \\ \cline{5-5} 
Sh 2-209           & 1 $\pm$ 0.2       & 1000                                     & 2022 & \cite{Yasui2023}                     \\ \cline{5-5} 
Tr 14              & 1.3 $\pm$ 0.1     & 8847                                     & 2012 & \cite{Hur2012,Ascenso2007}           \\ \cline{5-5} 
Wes 1              & 0.8 $\pm$ 0.1     & 7.3$\times 10^4$                                    & 2012 & \cite{Lim2013}                       \\ \hline
\end{tabular}
\end{table*}

\end{appendix}

%
%

\end{document}